\documentclass[a4paper,11pt]{article}
\pdfoutput=1 

\usepackage{jheppub} 

\usepackage[T1]{fontenc} 

\usepackage{amsmath}
\usepackage{amssymb}
\usepackage{graphicx}
\usepackage{xcolor}
\usepackage{multirow}
\usepackage{float}
\usepackage{makecell}
\usepackage{longtable}
\preprint{USTC-ICTS/PCFT-26-07}

\title{Chiral Integrable Boundary States of ABJM Spin Chain from Reflection Equations}

\author[a]{Yang~Liu,}
\author[b,1]{Nan~Bai,}
\author[b]{Mao-Zhong~Shao,}
\author[a, c, 1]{Jun-Bao~Wu\note{ Corresponding author.} }

\affiliation[a]{Center for Joint Quantum Studies and Department of Physics, School of Science, Tianjin University, 135 Yaguan Road, Tianjin 300350, P. R. China }
\affiliation[b]{Department of Physics, Guangxi Normal University, Guilin 541004, P. R. China}
\affiliation[c]{Peng Huanwu Center for Fundamental Theory, 96 Jinzhai Road, Hefei, Anhui 230026, P. R. China}

\emailAdd{liuyang\_2306@tju.edu.cn}
\emailAdd{bainan@mailbox.gxnu.edu.cn}
\emailAdd{mzshao@stu.gxnu.edu.cn}
\emailAdd{junbao.wu@tju.edu.cn}

\abstract{We develop a general framework for constructing $2n$-site chiral integrable matrix product states in Aharony-Bergman-Jafferis-Maldacena spin chain, based on reflection equations and the fusion procedure. For four-site chiral integrable product states, we propose their exact overlap formulas with  Bethe states. We also investigate the chiral integrable subspaces numerically.}

\begin{document} 
\maketitle
\flushbottom
\section{Introduction}
Matrix product states (MPS), as a key mathematical structure, appear in both the theoretical high energy physics and statistical physics over the past ten years. In particular, the integrable MPS have attracted considerable attention due to their ability to produce exact analytical results. The precise definition of integrable boundary MPS and their connection to the boundary integrability have been revealed in \cite{Piroli:2017sei,Pozsgay:2018dzs}, which extends an earlier work in two-dimensional integrable field theory \cite{Ghoshal:1993tm}.  Concretely, in the study of quantum quenches, the integrable boundary states (often termed as integrable initial states in the Wick-rotated picture), serve as the pre-quench ground states and play an important role in characterizing the post-quench dynamics by evaluating their overlaps with the Bethe eigenstates. There are vast investigations devoted to this direction, see, e.g., \cite{ Caux:2013ra,Wouters:2014,Pozsgay:2014,Mestyan:2017xyk,Piroli:2018ksf,Piroli:2018don,Rylands:2022gev,Rylands:2022naf}. In AdS/CFT holography, integrable MPS arise in the computation of correlation functions within various defect CFT setups. These include co-dimension one domain walls induced by intersecting probe branes, the dual giant graviton by a probe D3-brane, the extended objects such as Wilson and 't~Hooft lines, and vacuum condensates on the Coulomb branch. There are also a large amount of related works. A non-exhaustive summary includes references \cite{deLeeuw:2015hxa,Buhl-Mortensen:2015gfd,deLeeuw:2016umh,deLeeuw:2017dkd,Buhl-Mortensen:2017ind,DeLeeuw:2018cal,Jiang:2019xdz,
Jiang:2019zig,DeLeeuw:2019ohp,Komatsu:2020sup,Kristjansen:2020mhn,Kristjansen:2023ysz,Ivanovskiy:2024vel,Gombor:2024api,Kristjansen:2024map,deLeeuw:2024qki,Holguin:2025bfe,Chalabi:2025nbg,Coronado:2025xwk,Demjaha:2025axy,Gombor:2025qvk} for the $\mathcal{N}=4$ SYM theory in the AdS$_5$/CFT$_4$ duality  and references \cite{Yang:2021hrl,Kristjansen:2021abc,Gombor:2022aqj,Yang:2022dlk,Jiang:2023cdm,Wu:2024uix,Bai:2024qtg,Liu:2025uiu} in the parallel holographic cousin of Aharony-Bergman-Jafferis-Maldacena (ABJM) theory \cite{Aharony:2008ug} in the AdS$_4$/CFT$_3$ context, as well as the related investigations in the dual gravity side \cite{Linardopoulos:2022wol,Linardopoulos:2021rfq,Linardopoulos:2025ypq, Yang:2021kot}. In all known cases, the overlaps take compact closed forms, thus serving as a remarkable property of integrable boundary states. Using the algebraic Bethe ansatz (ABA), the exact overlap formulas for general two-site product states in any $\mathfrak{gl}_N$ representation have been derived and proven in a series of works \cite{Gombor:2021uxz,Gombor:2021hmj,Gombor:2023bez}, and have been recently extended to integrable MPS overlaps of rational spin chains with arbitrary representations \cite{Gombor:2024iix,Gombor:2025wvu}.

Integrable boundary states can be further classified by their chirality as proposed in \cite{Gombor:2020kgu}.  For general high rank spin chains, the notion of chirality is introduced by the specific paring structure among different types of the Bethe roots: an integrable boundary state and its overlap with the Bethe state are called chiral if the non-vanishing overlap enforces the Bethe roots to be paired within the same type; otherwise, they are achiral, indicating the paring mixes different types of the Bethe roots. While both chiral and achiral integrable boundary states have been extensively investigated in the context of $\rm{AdS_5/CFT_4}$ correspondence, studies in the parallel $\rm{AdS_4/CFT_3}$ setting have mainly focused on the achiral ones. Investigations on chiral integrable states were recently initiated  by two of the present authors \cite{Liu:2025uiu}.
Therein, a sufficient condition for chiral integrable states was derived by expanding the transfer matrix in a specific form  (such approach has previously been applied to the achiral cases in \cite{Yang:2022dlk,Wu:2024uix}), and from this condition, the chiral integrable states can be identified among the spin chain basis states. Nevertheless, this exploration is limited to the pure basis states and lacks a general construction for chiral integrable states.

One the other hand, a large class of integrable boundary states can be systematically constructed from boundary integrability \cite{Piroli:2017sei,Pozsgay:2018dzs}. Given a $K$-matrix solution to the reflection equation (RE), also known as the boundary Yang-Baxter equation (BYBE),  one can use it to define a two-site state as the building block. Then the tensor products of these blocks gives rise to a product state which is two-site translationally invariant and automatically satisfies the integrability condition due to the RE. If the $K$-matrix further obeys the square-root relation introduced in \cite{Pozsgay:2018dzs}, the two-site block factorizes into a tensor product of two identical one-site blocks, and the resulting MPS reduces to a standard one-site integrable MPS. 
Investigations on more general $2n$-site translationally invariant integrable MPS are rare in the literature. Besides the specific case of $4$-site integrable basis states presented in \cite{Liu:2025uiu}, a general construction remains largely absent.

In the present paper we address both issues mentioned above and make substantial progress in each direction. In the study, we rely on the fusion technique, which is a procedure for constructing novel solutions of Yang-Baxter equation (YBE) and RE from the fundamental $R$- and $K$-matrices \cite{Kulish:1981gi,Mezincescu:1991ke} (see also its recent application in ABJM theory \cite{Bai:2025mpw}). By analyzing various reflection equations, we identify those whose $K$-matrices allow for the construction of chiral integrable MPS. With this identification, a suitable $n$-fused $K$-matrix corresponds to a $2n$-site chiral integrable MPS. We also investigate operator-valued $K$-matrices, which correspond to reflection processes involving additional boundary degrees of freedom. From such $K$-matrices, we can construct integrable boundary MPS with bond dimension greater than one. Finally, with the chiral integrable MPS at hand, the next central task is to determine their overlap with the Bethe states. Given that our chiral integrable MPS are constructed from a broad class of $K$-matrix solutions, a natural question is whether their overlaps still admit a compact form. Our findings support this claim: the overlap is expressed by the ratio of two Gaudin-like determinants, multiplied by scalar factors including ones originating from the $K$-matrix.

The rest of the paper is organized as follows. In section \ref{sec:review}, we review the basic ingredients of integrable boundary states and integrable ABJM spin chain, and then define chiral integrable MPS of the ABJM spin chain, which are the central object of the present work. In section \ref{sec:construction}, we give concrete constructions of chiral integrable MPS that originate from $K$-matrix solutions of specific reflection equations. In section \ref{sec:overlap} and \ref{sec:subspace}, we propose an exact overlap formula for the 4-site chiral integrable MPS and also investigate the chiral integrable subspaces based on some numerical data of small site number. In section \ref{sec:conclusion}, we conclude and outline some potential directions for future research. Finally, in the appendix \ref{roots}, we provide the Bethe root data used in our numerical analysis.

\section{Integrable boundary states in ABJM theory}\label{sec:review}

This section is devoted to a brief overview of the integrable structure of the $SU(4)$ alternating spin chain relevant to the ABJM theory. We introduce the notation for the Hamiltonian, the $R$-matrices, and transfer matrices, and recall the Bethe ansatz description of the spectrum. Building on these elements, we then review the notion of integrable boundary states, which will serve as the defining property for the chiral integrable boundary states in the following sections.

At planar two-loop order, the dilatation operator acting on scalar sector in ABJM theory is mapped to the Hamiltonian of an integrable $SU(4)$ alternating spin chain \cite{Minahan:2008hf, Bak:2008cp}
\begin{align}
    \mathrm{H} = \frac{\lambda^2}{2} \sum_{l=1}^{2L} \left( 2 - 2\mathbb{P}_{l, l+2} + \mathbb{P}_{l, l+2}\mathbb{K}_{l, l+1}+\mathbb{K}_{l, l+1}\mathbb{P}_{l, l+2} \right), \label{H}
\end{align}
where $\mathbb{P}_{ij}$ and $\mathbb{K}_{ij}$ denote the permutation and trace operators, respectively, and
 $\lambda$ is the ’t Hooft coupling. 
 The spin chain consists of $2L$ sites, with odd (even) sites transforming in the fundamental ($\mathbf{4}$) (anti-fundamental ($\bar{\mathbf{4}}$) ) representation of $SU(4)$. The action of the permutation and trace operators is defined as
\begin{align}
\mathbb{P}\left | A \right \rangle \otimes \left | B \right \rangle  &= \left | B \right \rangle \otimes  \left | A \right \rangle,\quad 
\mathbb{P}\left | \bar{A}  \right \rangle \otimes \left | \bar{B}  \right \rangle  = \left | \bar{B}  \right \rangle \otimes  \left | \bar{A}  \right \rangle;\\
\mathbb{K}\left | A \right \rangle \otimes \left | \bar{B} \right \rangle&=\delta_{AB} \sum_{C=1}^{4} \left | C \right \rangle \otimes \left | \bar{C}  \right \rangle, \quad 
\mathbb{K}\left | \bar{A}  \right \rangle \otimes \left | B  \right \rangle  = \delta_{AB} \sum_{C =1}^{4} \left | \bar{C} \right \rangle \otimes \left | C \right \rangle, 
\end{align}
where $A,B,C=1,\ldots,4$, and barred indices label the anti-fundamental representation.
The operators $\mathbb{P}_{ij}$ and $\mathbb{K}_{ij}$ satisfy a number of algebraic relations
\begin{align}
\mathbb{P}_{ij}\mathbb{P}_{ij}&=\mathbb{I}_{ij},\quad \mathbb{P}_{ij}\mathbb{P}_{ik}=\mathbb{P}_{ik}\mathbb{P}_{kj}=\mathbb{P}_{jk}\mathbb{P}_{ij},\\
\mathbb{K}_{ij}\mathbb{K}_{ij}&=4\mathbb{K}_{ij},\quad \mathbb{K}_{ij}\mathbb{K}_{ik}=\mathbb{K}_{ij}\mathbb{P}_{jk}=\mathbb{P}_{jk}\mathbb{K}_{ik},\\
\text{tr}_i\mathbb{P}_{ij}&=\mathbb{I}_j,\quad \text{tr}_i\mathbb{K}_{ij}=\mathbb{I}_j.
\end{align}
It is also useful to employ the explicit index notation for these operators,
\begin{align}\label{eq:components}
\mathbb{I}_{a_1a_2}^{b_1b_2}=\delta_{a_1}^{b_1}\delta_{a_2}^{b_2},\quad \mathbb{P}_{a_1a_2}^{b_1b_2}=\delta_{a_1}^{b_2}\delta_{a_2}^{b_1},\quad \mathbb{K}_{a_1\bar{a}_2}^{b_1\bar{b}_2}=\delta^{{b}_1\bar{b}_2}\delta_{a_1\bar{a}_2},
\end{align}
which will be used in explicit computations latter.
 
Integrability of the alternating chain is encoded in four $R$-matrices,
\begin{align}
R_{0j}(u) &= u \mathbb{I} + \mathbb{P}_{0j}, \nonumber \\
R_{0\bar{j}}(u) &= -(u + 2) \mathbb{I} + \mathbb{K}_{0\bar{j}}, \nonumber \\
R_{\bar{0}j}(u) &= -(u + 2) \mathbb{I} + \mathbb{K}_{\bar{0}j}, \nonumber \\
R_{\bar{0}\bar{j}}(u) &= u \mathbb{I} + \mathbb{P}_{\bar{0}\bar{j}}, \label{R}
\end{align}
where the auxiliary space $0$ ($\bar{0}$) transforms in the fundamental (anti-fundamental) representation of $SU(4)$. 
For later clarity, we adopt the following shorthand notation, in which the four $R$-matrices introduced above are grouped into two classes, 
\begin{align}
R_{0j/\bar{0}\bar{j}}(u) & = u \mathbb{I} + \mathbb{P}_{0j/\bar{0}\bar{j}}, \\
\bar{R}_{0\bar{j}/\bar{0}j}(u) & = -(u + 2) \mathbb{I} + \mathbb{K}_{0\bar{j}/\bar{0}j}.
\end{align}
Labeling the sites of the chain as $1, \bar{1}, 2, \bar{2},\cdots, L,\bar{L}$, one introduces two monodromy matrices,
\begin{align}
    T_0(u) &= R_{01}(u) \bar{R}_{0\bar{1}}(u) \cdots R_{0L}(u) \bar{R}_{0\bar{L}}(u),\label{M0}\\
    \bar{T}_{\bar{0}}(u) &= \bar{R}_{\bar{0}1}(u) R_{\bar{0}\bar{1}}(u) \cdots \bar{R}_{\bar{0}L}(u) R_{\bar{0}\bar{L}}(u),\label{M0b}
\end{align}
whose traces over the auxiliary space define the transfer matrices,
\begin{align}
\tau(u) &= \text{tr}_0 T_0(u), \quad \bar{\tau}(u) = \text{tr}_{\bar{0}} \bar{T}_{\bar{0}}(u). \label{t}
\end{align}

The spectrum of the alternating spin chain can be constructed via the nested algebraic Bethe ansatz~\cite{Kulish:1983rd, Minahan:2008hf, Bak:2008cp}. The eigenstates can be obtained through the nested coordinate Bethe ansatz \cite{Yang:2021hrl}. A generic on-shell Bethe state is characterized by three sets of rapidities, $\mathbf{u}$, $\mathbf{w}$ and $\mathbf{v}$, satisfying the Bethe equations
\begin{align}
    \begin{gathered}\begin{aligned}1=e^{i \phi_{u_j}}:=\left(\frac{u_j+\frac{i}{2}}{u_j-\frac{i}{2}}\right)^L\prod_{\begin{subarray}{c}k=1\\k\neq j\end{subarray}}^{N_{\mathbf{u}}}S(u_j,u_k)\prod_{k=1}^{N_{\mathbf{w}}}\tilde{S}(u_j,w_k), \end{aligned}\\\begin{aligned}1=e^{i \phi_{w_j}}:=\prod_{\begin{subarray}{c}k=1\\k\neq j\end{subarray}}^{N_{\mathbf{w}}}S(w_j,w_k)\prod_{k=1}^{N_{\mathbf{u}}}\tilde{S}(w_j,u_k)\prod_{k=1}^{N_{\mathbf{v}}}\tilde{S}(w_j,v_k), \end{aligned}\\\begin{aligned}1=e^{i \phi_{v_j}}:=\left(\frac{v_j+\frac{i}{2}}{v_j-\frac{i}{2}}\right)^L\prod_{\begin{subarray}{c}k=1\\k\neq j\end{subarray}}^{N_\mathbf{v}}S(v_j,v_k)\prod_{k=1}^{N_\mathbf{w}}\tilde{S}(v_j,w_k),\end{aligned}\end{gathered}\label{bae}
\end{align}
where $N_u, N_w, N_v$ denote the numbers of rapidities in the sets $\mathbf{u} = \{u_1, \dots, u_{N_u}\}$, $\mathbf{w} = \{w_1, \dots, w_{N_w}\}$, $\mathbf{v} = \{v_1, \dots, v_{N_v}\}$, respectively, and with scattering factors
\begin{align}
S(u, v) &= \frac{u - v - i}{u - v + i}, \quad \tilde{S}(u, v) = \frac{u - v + \frac{i}{2}}{u - v - \frac{i}{2}}. \label{scatter}
\end{align}
$\phi_{u_j}, \phi_{w_j}, \phi_{v_j}$ are introduced for the later definition of the Gaudin matrix in subsection \ref{subsec:Gaudin}.
The corresponding Bethe states $\left | \bf{u}, \bf{v}, \bf{w}\right\rangle$ are built on the reference vacuum $\left |1\bar{4}\right\rangle^{\otimes L}$.
The eigenvalues of the transfer matrices $\tau(u)$ and $\bar{\tau}(u)$ on these states are given by
\begin{equation}\label{lmd}
\begin{aligned}
\Lambda(u)=&
(-u-2)^L(u+1)^L\prod_{i=1}^{N_u}\frac{u-iu_i-1/2}{u-iu_i+1/2}
\\
&+(-u)^L(u+1)^L\prod_{i=1}^{N_v}\frac{u-iv_i+5/2}{u-iv_i+3/2}\\
&+(-u)^L(u+2)^L\prod_{i=1}^{N_u}\frac{u-iu_i+3/2}{u-iu_i+1/2}\prod_{i=1}^{N_w}\frac{u-iw_i}{u-iw_i+1}\\
&+(-u)^L(u+2)^L\prod_{i=1}^{N_v}\frac{u-iv_i+1/2}{u-iv_i+3/2}\prod_{i=1}^{N_w}\frac{u-iw_i+2}{u-iw_i+1}, 
\end{aligned}
\end{equation}
and
\begin{equation}\label{lmdb}
\begin{aligned}
    \bar{\Lambda}(u)=&(-u)^L(u+1)^L\prod_{i=1}^{N_u}\frac{u-iu_i+5/2}{u-iu_i+3/2}\\
    &+(-u-2)^L(u+1)^L\prod_{i=1}^{N_v}\frac{u-iv_i-1/2}{u-iv_i+1/2}\\
    &+(-u)^L(u+2)^L\prod_{i=1}^{N_u}\frac{u-iu_i+1/2}{u-iu_i+3/2}\prod_{i=1}^{N_w}\frac{u-iw_i+2}{u-iw_i+1}\\
    &+(-u)^L(u+2)^L\prod_{i=1}^{N_v}\frac{u-iv_i+3/2}{u-iv_i+1/2}\prod_{i=1}^{N_w}\frac{u-iw_i}{u-iw_i+1}, 
\end{aligned}
\end{equation}
respectively.

Integrable boundary states are characterized by their annihilation by all odd conserved charges, a condition that can be formulated in terms of transfer matrices. In the ABJM alternating spin chain, this leads to two distinct classes of integrable boundary conditions. Chiral integrable states satisfy the untwisted condition
\begin{align}
\tau(u)\left |\mathcal{B} \right\rangle=\Pi \tau(u)\Pi \left |\mathcal{B} \right\rangle\, \Leftrightarrow \,
\bar{\tau}(u)\left |\mathcal{B} \right\rangle=\Pi \bar{\tau}(u)\Pi \left |\mathcal{B} \right\rangle\, \Leftrightarrow \,\tau (u) \left | \mathcal{B}  \right \rangle=\bar{\tau } (-u-2) \left | \mathcal{B}  \right \rangle, \label{untwisted condition} 
\end{align}
with $\Pi$ being the reflection operator
\begin{align}
    \Pi \left |A_1 A_{\bar{1}} \cdots A_L A_{\bar{L}}\right \rangle
    =\left| A_{\bar{L}} A_L \cdots A_{\bar{1}} A_1 \right\rangle.
\end{align}
Achiral integrable states instead obey the twisted condition
\begin{align}
\tau(u)\left |\mathcal{B} \right\rangle=\Pi \bar{\tau}(u)\Pi \left |\mathcal{B} \right\rangle\, \Leftrightarrow \,
\bar{\tau}(u)\left |\mathcal{B} \right\rangle=\Pi \tau(u)\Pi \left |\mathcal{B} \right\rangle \,\Leftrightarrow \,\tau (u) \left | \mathcal{B}  \right \rangle=\tau(-u-2) \left | \mathcal{B}  \right \rangle.
\end{align}
In the remainder of this work, we will focus exclusively on chiral integrable boundary states.

The transfer-matrix characterization of chiral integrable boundary states reviewed in this section will be our main tool in what follows. In particular, it allows for a direct construction of integrable boundary states of the alternating spin chain from solutions to RE and we will make this connection explicit in the next section.

\section{Chiral integrable boundary states from reflection $K$-matrices} \label{sec:construction}

In this section, we address a systematic construction of integrable boundary states from RE. The close relationship between the two has been extensively explored in \cite{Piroli:2017sei,Pozsgay:2018dzs,Piroli:2018ksf,Piroli:2018don}. 
Roughly speaking, the connection between them is twofold: First, the partition function of the lattice field theory can be expressed in terms of the quantum transfer matrix (QTM). When the integrable state is constructed from the $K$-matrix, which is the solution of the RE, the corresponding QTM becomes proportional to an open spin chain transfer matrix. Second, the boundary state constructed from the $K$-matrix automatically satisfies the integrable boundary condition and thus becomes an integrable boundary state. In the following, we will give a detailed investigation on the construction of chiral integrable MPS from various RE.

\subsection{SP- and SNP-type reflection equations}
For the ABJM open spin chain, in addition to the four $R$-matrices introduced in section \ref{sec:review} which describe the bulk scattering, we also need to introduce boundary reflection $K$-matrices to describe how  solitons or anti-solitons reflect off the boundary. The corresponding integrability is guaranteed by the RE \cite{Sklyanin:1988yz}~\footnote{The integrable ABJM open chain~\cite{Chen:2018sbp} from the maximal giant graviton was solved by ABA in  \cite{Bai:2019soy}. Such results are still lacking for the open chain from flavored ABJM theory~\cite{Bai:2017jpe}, though coordinate Bethe ansatz gave strong evidence that this chain is integrable as well.}.  Boundary reflections can be distinguished into two types. The first is soliton-preserving (SP) reflection, where a soliton or anti-soliton reflects without changing its type. The second is soliton non-preserving (SNP) reflection, in which a soliton reflects as an anti-soliton and vice versa. The SNP case was first investigated for the $SU(3)$-invariant model in \cite{Doikou:2000yw} and later generalized to $SU(N)$ and $SU(M|N)$ case in \cite{Arnaudon:2004sd}.

\begin{figure}[h]
 \begin{center}
   \includegraphics[width=0.7\linewidth]{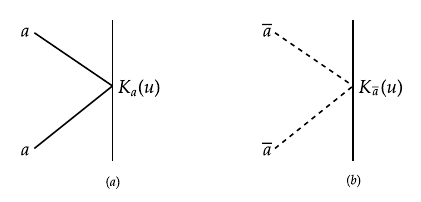}
 \end{center}
\caption{Two SP-type reflection $K$-matrices: (a) soliton to soliton reflection; (b) anti-soliton to anti-soliton reflection.  \label{fig:SPK}}
\end{figure}

We now formulate the SP- and SNP-type RE in $SU(4)$ ABJM spin chain. For the SP-type reflection, we introduce two $K$-matrices, $K_a(u)$ and $K_{\bar{a}}(u)$ shown in Fig.~\ref{fig:SPK}\footnote{The subscripts $a$ and $\bar{a}$ denote the $SU(4)$ fundamental and anti-fundamental representations carried by SP-type $K$-matrices.}, which describe soliton to soliton reflection and anti-soliton to anti-soliton reflection, respectively. 

\begin{figure}[h]
 \begin{center}
   \includegraphics[width=0.9\linewidth]{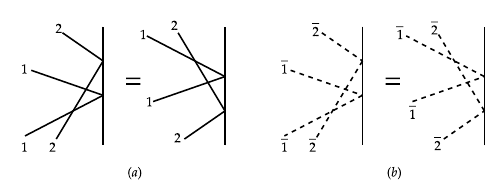}
 \end{center}
\caption{Two SP-type reflections. \label{fig:SPRE}}
\end{figure}

Two SP-type RE \footnote{In the following, for notational simplicity, we no longer use the bar to distinguish whether the two spaces on which the $R$-matrix acts are in the fundamental or the anti-fundamental representation of $SU(4)$, as this will be clear from the context.} are given below, 
\begin{eqnarray}
&&R_{12}(u-v)K_1(u)R_{21}(u+v)K_2(v)=K_2(v)R_{12}(u+v)K_1(u)R_{21}(u-v),\\
&&R_{12}(u-v)K_{\bar{1}}(u)R_{21}(u+v)K_{\bar{2}}(v)=K_{\bar{2}}(v)R_{12}(u+v)K_{\bar{1}}(u)R_{21}(u-v),
\end{eqnarray}
which correspond to the reflection processes shown in Fig.~\ref{fig:SPRE}.

\begin{figure}[h]
 \begin{center}
   \includegraphics[width=0.7\linewidth]{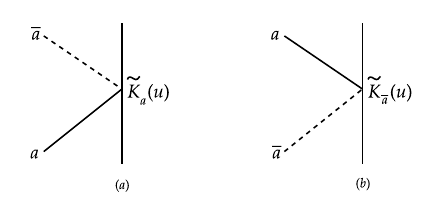}
 \end{center}
\caption{Two SNP-type reflection $K$-matrices: (a) soliton to anti-soliton reflection; (b) anti-soliton to soliton reflection.  \label{fig:SNPK}}
\end{figure}

The SNP-type reflection is similarly characterized by two $K$-matrices, $\tilde{K}_a(u)$ and $\tilde{K}_{\bar{a}}(u)$ depicted in Fig.~\ref{fig:SNPK}, 
 describing soliton to anti-soliton reflection and anti-soliton to soliton reflection, respectively. 

The SNP-type RE (also known as twisted RE) are given as
\begin{eqnarray}
&&R_{12}(u-v)\tilde{K}_{1}(u)\bar{R}_{21}(u+v)\tilde{K}_{2}(v)=\tilde{K}_{2}(v)\bar{R}_{12}(u+v)\tilde{K}_{1}(u) R_{21}(u-v),\\
&&R_{12}(u-v)\tilde{K}_{\bar{1}}(u)\bar{R}_{21}(u+v)\tilde{K}_{\bar{2}}(v)=\tilde{K}_{\bar{2}}(v)\bar{R}_{12}(u+v)\tilde{K}_{\bar{1}}(u) R_{21}(u-v),\label{eq:basicre3}
\end{eqnarray}
\begin{figure}[h]
 \begin{center}
   \includegraphics[width=0.9\linewidth]{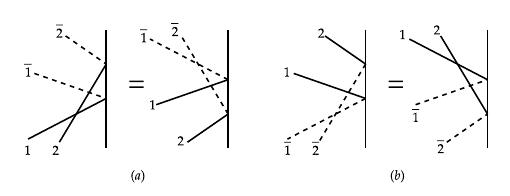}
 \end{center}
\caption{Two SNP-type reflections. \label{fig:SNPRE}}
\end{figure} 
which correspond to the reflection processes shown in Fig.~\ref{fig:SNPRE}.

We may also consider the scattering involving two distinct SNP (or two distinct SP) boundary reflections. These will be treated carefully in subsequent subsections, where the corresponding RE prove crucial for constructing chiral integrable MPS.

\subsection{Two-site integrable MPS}\label{subsec:two-site}
\begin{figure}[h]
 \begin{center}
   \includegraphics[width=0.5\linewidth]{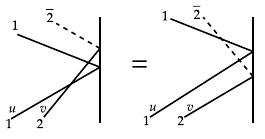}
 \end{center}
\caption{SP- and SNP-mixed reflection process. \label{fig:mixre}}
\end{figure}

In this section, we focus on constructing chiral integrable MPS with two-site translation invariance. To obtain the correct untwisted integrability condition $\eqref{untwisted condition}$, we consider a mixed boundary scattering process involving  both SP and SNP reflections, as shown in Fig.~\ref{fig:mixre}, whose RE is given by
\begin{equation}\label{eq:mixre}
R_{12}(u-v)K_1(u)R_{21}(u+v)\tilde{K}_2(v)=\tilde{K}_2(v) \bar{R}_{12}(u+v)K_1(u)\bar{R}_{21}(u-v).
\end{equation}
Rewriting the above equation (\ref{eq:mixre}) in terms of states, we have
\begin{equation}
R_{01}(u)\bar{R}_{02}(u)|\phi(-1)\rangle_{12}\otimes |\tilde{\phi}(-u-1)\rangle_{00'}
=\bar{R}_{0'2}(u)R_{0'1}(u)|\tilde{\phi}(-u-1)\rangle_{00'}\otimes|\phi(-1)\rangle_{12},
\end{equation}
where the two-site states are defined as
\begin{equation}
|\phi(u)\rangle=\left[K_1(u)\right]^i_j|i,j\rangle,\quad |\tilde{\phi}(u)\rangle=\left[\tilde{K}_1(u)\right]^i_j|i,j\rangle.
\end{equation}
Thus, the existence of an invertible solution $\tilde{K}_a$ for (\ref{eq:mixre}) implies that the tensor-product state
\begin{equation}
|\psi\rangle := |\phi(-1)\rangle_{12} \otimes \cdots \otimes |\phi(-1)\rangle_{2L-1,2L},
\end{equation}
satisfies
\begin{equation}\label{eq:2siteic}
\tau (u)|\psi\rangle = \Pi \tau (u) \Pi |\psi\rangle ,
\end{equation}
therefore becoming a chiral integrable MPS\footnote{Here we regard the integrable product state constructed from the c-number solution of the reflection equation as an MPS with bond dimension one. The generalization to MPS with a generic bond dimension is straightforward: one takes the tensor product of operator-valued $K$-matrix solutions and then trace over the internal bond space. Construction details are demonstrated  in section \ref{Sec:nMPS} with a concrete integrable MPS of bond dimension four. }. However, our numerical results indicate that the constant solutions for $K$ and $\tilde{K}$ in (\ref{eq:mixre}) are trivial, requiring that one of them must be the zero matrix. Based on this result, we further conjecture that there is no nontrivial invertible $\tilde{K}(u)$-matrix solution for (\ref{eq:mixre}). A heuristic but not rigorous argument is as follows. From Eq.~(\ref{eq:mixre}), we observe that $\tilde{K}(v)$ can be seen as an intertwiner between the following two operators:
\begin{equation}
\mathcal{O}_A=R_{12}(u-v)K_1(u)R_{21}(u+v),\quad \mathcal{O}_B=\bar{R}_{12}(u+v)K_1(u)\bar{R}_{21}(u-v).
\end{equation}
Assume that $\mathcal{O}_A$ and $\mathcal{O}_B$ transform under the $\mathbf{4} \otimes \mathbf{4}$ and $\mathbf{4} \otimes \bar{\mathbf{4}}$ representations of $SU(4)$, respectively. Since the tensor product representation spaces decompose as
\begin{equation}
\mathbf{4}\otimes\mathbf{4}=\mathbf{6}_\mathrm{A}\oplus \mathbf{10}_\mathrm{S},\quad \mathbf{4}\otimes \bar{\mathbf{4}}=\mathbf{1}\oplus \mathbf{15}_{\rm{adj}},
\end{equation}
they share no isomorphic irreducible sub-representations. Thus, the intertwiner $\tilde{K}(v)$ cannot be invertible. In the section \ref{sec:subspace}, we will directly solve $\eqref{untwisted condition}$ to search for possible 2-site chiral MPS with some small site numbers.

\subsection{Four-site integrable MPS}
\begin{figure}[h]
 \begin{center}
   \includegraphics[width=0.6\linewidth]{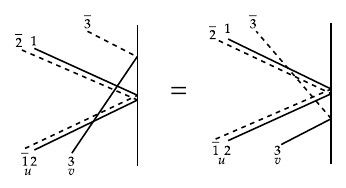}
 \end{center}
\caption{Boundary reflection corresponding to four-site integrable state. \label{fig:fusedsc}}
\end{figure}

\begin{figure}[h]
 \begin{center}
   \includegraphics[width=0.9\linewidth]{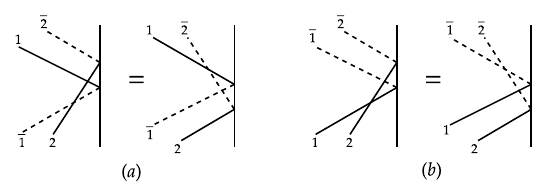}
 \end{center}
\caption{Two fundamental SNP-type boundary reflections. \label{fig:basicre}}
\end{figure}

In this section, we construct the four-site invariant integrable MPS which satisfies the chiral integrability condition $\eqref{untwisted condition}$. There are several ways to realize such integrable states from concrete boundary scattering processes. One such construction is to consider a 2-2 scattering between a single soliton and a soliton-anti-soliton bound state. Integrability requires two equivalent boundary scattering processes as depicted in Fig.~\ref{fig:fusedsc}, which corresponds to the following fused RE:
\begin{equation}\label{eq:fusedre}
R_{(\bar{1}2),3}(u-v)\tilde{K}_{(\bar{1}2)}(u)R_{3,(\bar{2}1)}(u+v)\tilde{K}_3(v)=
\tilde{K}_3(v)R_{(\bar{1}2),\bar{3}}(u+v)\tilde{K}_{(\bar{1}2)}(u)R_{\bar{3},(\bar{2}1)}(u-v),
\end{equation}
where the four fused $R$-matrices are given by
\begin{eqnarray}
&&R_{(\bar{1}2),3}(u)=R_{23}(u)\bar{R}_{13}(u),\\
&&R_{3,(\bar{2}1)}(u)=\bar{R}_{32}(u)R_{31}(u),\\
&&R_{(\bar{1}2),\bar{3}}(u)=\bar{R}_{23}(u)R_{13}(u),\\
&&R_{\bar{3},(\bar{2}1)}(u)=R_{32}(u)\bar{R}_{31}(u),
\end{eqnarray}
and the fused $K$-matrix $\tilde{K}_{(\bar{1}2)}(u)$ is composed of two fundamental SNP reflections,
\begin{equation}\label{eq:fusedK}
\tilde{K}_{(\bar{1}2)}(u)=\tilde{K}_2(u)R_{12}(2u)\tilde{K}_{\bar{1}}(u).
\end{equation}
The fused RE (\ref{eq:fusedre}) relies on the two basic boundary reflections, as illustrated in Fig.\ref{fig:basicre}.
We thus require the following two fundamental SNP RE for $\tilde{K}_a$ and $\tilde{K}_{\bar{a}}$-matrices:
\begin{eqnarray}
&&\bar{R}_{12}(u-v)\tilde{K}_{\bar{1}}(u)R_{21}(u+v)\tilde{K}_2(v)=\tilde{K}_2(v)R_{12}(u+v)\tilde{K}_{\bar{1}}(u)\bar{R}_{21}(u-v),\label{eq:basicre1}\\
&&R_{12}(u-v)\tilde{K}_1(u)\bar{R}_{21}(u+v)\tilde{K}_2(v)=\tilde{K}_2(v)\bar{R}_{12}(u+v)\tilde{K}_1(u) R_{21}(u-v).\label{eq:basicre2}
\end{eqnarray}
Using the fused $K$-matrix (\ref{eq:fusedK}), we construct the four-site block
\begin{equation}
|\Phi(u)\rangle_{1234}=\left[\tilde{K}_{(\bar{1}2)}(u)\right]^{i_1,i_2}_{j_1,j_2}|i_2,i_1,j_2,j_1\rangle.
\end{equation}
We also define the two-site block $|\phi(u)\rangle=[\tilde{K}_1(u)]^i_j|i,j\rangle$. In terms of these states, the fused RE (\ref{eq:fusedre}) can be recast in the form:
\begin{equation}\label{eq:state}
\begin{split}
&{R}_{01}(u-v)\bar{R}_{02}(u-v){R}_{03}(-u-v-2)\bar{R}_{04}(-u-v-2)|\Phi(u)\rangle_{1234}\otimes |\phi(v)\rangle_{00'}\\
=&\bar{R}_{0'4}(u-v){R}_{0'3}(u-v)\bar{R}_{0'2}(-u-v-2){R}_{0'1}(-u-v-2)|\phi(v)\rangle_{00'}\otimes|\Phi(u)\rangle_{1234}.
\end{split}
\end{equation}
For a spin chain of length $2L$, we define the following tensor product state in terms of the four-site blocks: 
\begin{equation}\label{eq:4sitestate}
|\Psi\rangle=|\Phi(-1)\rangle_{1234}\otimes |\Phi(-1)\rangle_{5678}\otimes\cdots\otimes |\Phi(-1)\rangle_{2L-3,2L-2,2L-1,2L}.
\end{equation}
Thus, given any two-site state $|\phi(v)\rangle$ constructed from an invertible $\tilde{K}_a$-matrix, state equation (\ref{eq:state}) leads to
\begin{equation}\label{eq:4siteic}
\tau (u)|\Psi\rangle = \Pi \tau (u) \Pi |\Psi\rangle ,
\end{equation}
showing that $|\Psi\rangle$ is a four-site invariant (by construction) chiral integrable MPS.
\subsubsection{More general constructions}
\begin{figure}[h]
 \begin{center}
   \includegraphics[width=0.6\linewidth]{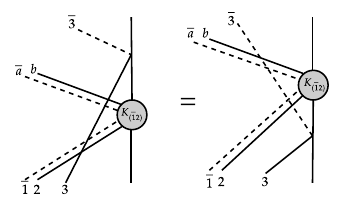}
 \end{center}
\caption{General boundary scattering for four-site chiral MPS. \label{fig:generalsc}}
\end{figure}

\begin{figure}[h]
 \begin{center}
   \includegraphics[width=0.9\linewidth]{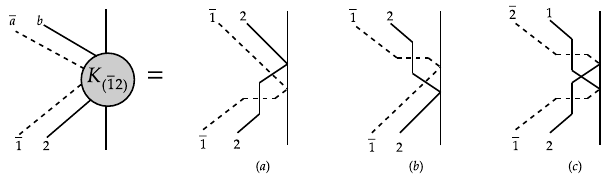}
 \end{center}
\caption{Three concrete constructions of fused $K$-matrix. \label{fig:fusedKs}}
\end{figure}

In the construction above, the fused $K$-matrix is built from two elementary SNP reflections. More generally, it can be realized through any composite scattering process that obeys both the YBE and the RE. The boundary scattering process involving such a general fused $K$-matrix is depicted in Fig.~\ref{fig:generalsc}, and the corresponding RE is given by
\begin{equation}\label{eq:2fused}
\begin{split}
R_{(\bar{1}2),3}(u-v){K}_{(\bar{1}2)}(u)R_{3,(\bar{a}b)}(u+v)\tilde{K}_3(v)=
\tilde{K}_3(v)R_{(\bar{1}2),\bar{3}}(u+v){K}_{(\bar{1}2)}(u)R_{\bar{3},(\bar{a}b)}(u-v),
\end{split}
\end{equation}
where $(\bar{a}b)=(\bar{2}1)$ or $(\bar{a}b)=(\bar{1}2)$. 
This equation can be rewritten as a boundary state equation that is identical in form to Eq.~(\ref{eq:state}), with the four-site block defined by:
\begin{eqnarray}
|\Phi(u)\rangle_{1234}=\left\{
\begin{array}{cc}
\left[{K}_{(\bar{1}2)}(u)\right]^{i_1,i_2}_{j_1,j_2}|i_2,i_1,j_1,j_2\rangle,\quad (\bar{a}b)=(\bar{1}2)\\
\left[{K}_{(\bar{1}2)}(u)\right]^{i_1,i_2}_{j_1,j_2}|i_2,i_1,j_2,j_1\rangle,\quad (\bar{a}b)=(\bar{2}1)
\end{array}
\right.
\end{eqnarray}
Here we give some concrete constructions for the fused ${K}_{(\bar{1}2)}(u)$ matrix: 
\begin{itemize}
  \item [(a)] The fused reflection consists of two SP-type fundamental reflections and one bulk scattering at the beginning, as depicted in Fig.~\ref{fig:fusedKs}(a). The corresponding explicit expression reads
      \begin{equation}
      K_{\bar{1}2}(u)=\bar{R}_{12}(0)K_{\bar{1}}(u)\bar{R}_{21}(2u)K_2(u).
      \end{equation}
  \item [(b)] The fused reflection includes two SP-type fundamental reflections together with a subsequent bulk scattering after the two boundary reflections. This process is shown in Fig.~\ref{fig:fusedKs}(b) and is given explicitly by
      \begin{equation}
       K_{\bar{1}2}(u)=K_2(u)\bar{R}_{12}(2u)K_{\bar{1}}(u)\bar{R}_{21}(0).
      \end{equation}
  \item [(c)] The fused reflection involves two SNP-type fundamental reflections and two bulk scattering processes as shown in Fig.~\ref{fig:fusedKs}(c). The exact expression is
      \begin{equation}
      K_{\bar{1}2}(u)=\bar{R}_{12}(0)\tilde{K}_{\bar{1}}(u)R_{21}(2u)\tilde{K}_2(u)\bar{R}_{12}(0).  
      \end{equation}
\end{itemize}

\subsubsection{$K$-matrix solutions from SNP-type reflection equation}
In this section, we investigate the explicit form of the fused $K$-matrix, from which concrete integrable MPS are obtained. To be specific, we adopt the construction given by Eq.~(\ref{eq:fusedK}), which represents the simplest setup that yields chiral integrable states. In this construction, the fused $K$-matrix consists of two SNP-type fundamental $K$-matrices $\tilde{K}_a$ and $\tilde{K}_{\bar{a}}$. The solution $\tilde{K}_a$ of the SNP RE (\ref{eq:basicre2}) has been investigated in detail in \cite{Arnaudon:2004sd}. In that work, a newly defined transpose $\rm{``t"}$ was employed, which is related to the conventional transpose $\rm{``T"}$ by relation: $A^t=V^{-1}A^T V$, where $V={\rm{antidiag}}(1,1,\cdots,1)$. This modified transpose is used to define the $R$-matrix between soliton and anti-soliton and also appears in the dual RE. In the present work, however, we use the conventional transpose throughout. It is therefore necessary to revisit the SNP $K$-matrix solution in (\ref{eq:basicre2}), which is given below using a slightly different approach from that in \cite{Arnaudon:2004sd}.

We begin by substituting the explicit expressions of $R$- and $\bar{R}$-matrices into (\ref{eq:basicre2}), which gives
\begin{equation}
\begin{split}
(u-v)\tilde{K}_{1}(u)\mathbb{K}_{12}\tilde{K}_2(v)+(-u-v-2)\mathbb{P}_{12}\tilde{K}_1(u)\tilde{K}_2(v)+\mathbb{P}_{12}\tilde{K}_1(u)\mathbb{K}_{12}\tilde{K}_2(v)\\
=(u-v)\tilde{K}_{2}(v)\mathbb{K}_{12}\tilde{K}_1(u)+(-u-v-2)\tilde{K}_2(v)\tilde{K}_{1}(u)\mathbb{P}_{12}+\tilde{K}_2(v)\mathbb{K}_{12}\tilde{K}_1(u)\mathbb{P}_{12}.
\end{split}
\end{equation}
Then we extract the symmetric part of the above equation under the exchange of the spectral parameters $u$ and $v$ (or under the exchange of the spaces $1\leftrightarrow2$) to get
\begin{equation}
\tilde{K}_{1}(u)\mathbb{K}_{12}\tilde{K}_2(v)+\tilde{K}_{2}(u)\mathbb{K}_{12}\tilde{K}_1(v)
=\tilde{K}_{2}(v)\mathbb{K}_{12}\tilde{K}_1(u)+\tilde{K}_{1}(v)\mathbb{K}_{12}\tilde{K}_2(u).
\end{equation}
To proceed, we take the partial transpose in space $V_1$ and multiply by $\mathbb{P}_{12}$ from the left, which gives
\begin{equation}\label{eq:transpose1}
\tilde{K}^T_1(u)\tilde{K}_2(v)+\tilde{K}_1(u)\tilde{K}^T_2(v)=\tilde{K}_1(v)\tilde{K}^T_2(u)+\tilde{K}_1^T(v)\tilde{K}_2(u).
\end{equation}
Transposing again on space $V_1$, we have
\begin{equation}\label{eq:transpose2}
\tilde{K}_1(u)\tilde{K}_2(v)+\tilde{K}^T_1(u)\tilde{K}^T_2(v)=\tilde{K}^T_1(v)\tilde{K}^T_2(u)+\tilde{K}_1(v)\tilde{K}_2(u).
\end{equation}
Combining eqs. (\ref{eq:transpose1}) and (\ref{eq:transpose2}), we obtain
\begin{eqnarray}
&&\left[\tilde{K}^T_1(u)+\tilde{K}_1(u)\right]\left[\tilde{K}_2(v)+\tilde{K}^T_2(v)\right]=
\left[\tilde{K}_1(v)+\tilde{K}_1^T(v)\right]\left[\tilde{K}^T_2(u)+\tilde{K}_2(u)\right]\label{eq:combine1}\\
&&\left[\tilde{K}^T_1(u)-\tilde{K}_1(u)\right]\left[\tilde{K}_2(v)-\tilde{K}^T_2(v)\right]=
\left[\tilde{K}_1(v)-\tilde{K}_1^T(v)\right]\left[\tilde{K}^T_2(u)-\tilde{K}_2(u)\right]\label{eq:combine2}.
\end{eqnarray}
If we define
\begin{equation}
S(u)=\tilde{K}(u)+\tilde{K}^T(u),\quad A(u)=\tilde{K}(u)-\tilde{K}^T(u),
\end{equation}
then eqs. (\ref{eq:combine1}) and (\ref{eq:combine2}) become
\begin{equation}\label{eq:SA}
S_1(u)S_2(v)=S_1(v)S_2(u),\quad A_1(u)A_2(v)=A_1(v)A_2(u).
\end{equation}
Thus we find the possible solutions of (\ref{eq:SA}) take the following forms:
\begin{equation}
S(u)=p(u) S\quad A(u)=q(u) A,
\end{equation}
where $p(u)$ and $q(u)$ are two arbitrary scalar functions, $S$ and $A$ are arbitrary constant symmetric and antisymmetric matrices, respectively: $S^T=S,\,A^T=-A$. By restricting $\tilde{K}_a(u)$ to the form $p(u)S+q(u)A$ and plugging this ansatz back into (\ref{eq:basicre2}), we obtain, up to an overall scalar factor,
\begin{equation}\label{eq:newK}
\tilde{K}_a(u)=(1+2u)S+A,
\end{equation}
where $S$ and $A$ satisfy certain constraints. To our knowledge, this solution (\ref{eq:newK}) for the SNP-type RE is new to the literature. However, if $\tilde{K}_a(u)$ is required to be invertible, we find that the possible solutions reduce to arbitrary constant symmetric or anti-symmetric matrices as given in \cite{Arnaudon:2004sd}.

As also noted in \cite{Arnaudon:2004sd}, the compatibility condition between the RE allows the solution $\tilde{K}_{\bar{a}}(u)$ to be taken as the inverse of $\tilde{K}_{a}$, i.e., $\tilde{K}_{\bar{a}} \propto \bigl(\tilde{K}_{a}\bigr)^{-1}$. Indeed, for invertible $\tilde{K}_{\bar{a}}(u)$, Eq.~(\ref{eq:basicre3}) can be transformed into
 \begin{equation}
 \bar{R}_{12}(u+v)\tilde{K}_{\bar{1}}(u)R_{21}(u-v)\tilde{K}^{-1}_{\bar{2}}(v)=
 \tilde{K}^{-1}_{\bar{2}}(v)\bar{R}_{12}(u-v)\tilde{K}_{\bar{1}}(u)\bar{R}_{21}(u+v).
 \end{equation}
Comparing this with Eq.~(\ref{eq:basicre1}) gives $\tilde{K}_{\bar{a}}(u)\propto \tilde{K}^{-1}_{a}(-u)$. Moreover, since the invertible $\tilde{K}_a$ is proportional to a constant matrix, we simply choose $u$-independent $\tilde{K}_{\bar{a}} = \tilde{K}_{a}^{-1}$ in the following calculations.

\subsection{Integrable MPS from dressing}\label{Sec:nMPS}
\begin{figure}[h]
 \begin{center}
   \includegraphics[width=0.6\linewidth]{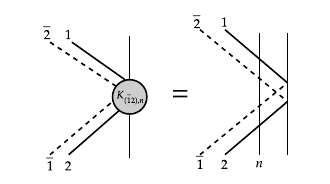}
 \end{center}
\caption{Fused $K$-matrix by dressing with $R$-matrices. \label{fig:dressedK}}
\end{figure}

So far, we have focused on MPS with bond dimension one. More general MPS with higher bond dimensions can be obtained by acting with the transfer matrix\cite{Piroli:2017sei,Pozsgay:2018dzs,DeLeeuw:2019ohp,Gombor:2024zru}. In this section, we will investigate the construction of such general four-site integrable MPS for ABJM alternating spin chain. From the viewpoint of the RE, MPS with bond dimension $>1$ correspond to the operator-valued $K$-matrices with internal degrees of freedom. Such operator-valued $K$-matrices can be obtained by dressing the ordinary c-number $K$-matrices with  $R$-matrices. To be concrete, we first consider dressing the fused $K$-matrices in (\ref{eq:fusedK}) by introducing an internal space carrying $SU(4)$ fundamental representation, as shown in Fig.~\ref{fig:dressedK}. 

The related fused $K$-matrix after dressing is given by
\begin{equation}
K_{(\bar{1}2),n}(u)=R_{2n}(u)\bar{R}_{1n}(u)\tilde{K}_{(\bar{1}2)}(u)\bar{R}_{n2}(u)R_{n1}(u),
\end{equation}
where $\tilde{K}_{(\bar{1}2)}(u)$ is defined in (\ref{eq:fusedK}), and the fused RE is
\begin{equation}\label{eq:dressedre}
R_{(\bar{1}2),3}(u-v){K}_{(\bar{1}2),n}(u)R_{3,(\bar{2}1)}(u+v)\tilde{K}_{3n}(v)=
\tilde{K}_{3n}(v)R_{(\bar{1}2),\bar{3}}(u+v){K}_{(\bar{1}2),n}(u)R_{\bar{3},(\bar{2}1)}(u-v),
\end{equation}
with
\begin{equation}
\tilde{K}_{3n}(u)=R_{3n}(u)\tilde{K}_3(u)\bar{R}_{n3}(u).
\end{equation}
Now let us decompose the $R$-matrices as
\begin{equation}
R_{0n}(u)=E_{ab}\otimes\mathcal{L}_{ab}(u),\quad \bar{R}_{0n}=E_{ab}\otimes \bar{\mathcal{L}}_{ab}(u),
\end{equation}
where $\{E_{ab},a,b=1,\cdots,4\}$ are the standard basis matrices and the exact forms of two operators $\mathcal{L}_{ab}(u)$ and $\bar{\mathcal{L}}_{ab}(u)$ can be easily extracted from the concrete expression of the related $R$-matrices. We also expand the fused $\tilde{K}_{(\bar{1}2)}(u)$ matrix as 
\begin{equation}
\tilde{K}_{(\bar{1}2)}(u)=\tilde{K}^{a_1a_2}_{b_1b_2}(u)E_{a_1b_1}\otimes E_{a_2b_2}\in {\rm{End}}(V_1\otimes V_2).
\end{equation}
Then the dressed fused $K_{(\bar{1}2),n}(u)$ matrix can be similarly expressed as
\begin{equation}
K_{(\bar{1}2),n}(u)=\left[K^D(u)\right]^{ab}_{cd}E_{ac}\otimes E_{bd}\in {\rm{End}}(V_1\otimes V_2\otimes V_n),
\end{equation}
with operator-valued components being
\begin{equation}
\left[K^D(u)\right]^{ab}_{cd}=\tilde{K}^{\alpha\beta}_{\gamma\delta}(u)\mathcal{L}_{b\beta}(u)\bar{\mathcal{L}}_{a\alpha}(u)\bar{\mathcal{L}}_{\delta d}(u)\mathcal{L}_{\gamma c}(u)\in {\rm{End}}(V_n).
\end{equation}
Now we define the dressed four-site block
\begin{equation}
|\Phi^D(u)\rangle=\left[K^D(u)\right]^{ab}_{cd}|b,a,d,c\rangle\in {\rm{End}}(V_n)\otimes V^{\otimes 4},
\end{equation}
from which we construct four-site integrable MPS with bond dimension four,
\begin{equation}
|\Psi^D\rangle=\text{tr}_{V_n}|\Phi^D(-1)\rangle_{1234}\otimes\cdots\otimes |\Phi^D(-1)\rangle_{2L-3,2L-2,2L-1,2L}\in V^{\otimes 2L}.
\end{equation}
If we define a new transfer matrix $t^D(u)$ for this length $2L$ spin chain as follows
\begin{equation}
t^D(u)=\text{tr}_{V_n}R_{1n}(u)\bar{R}_{2n}(u)\bar{R}_{3n}^{t_3}(u)R_{4n}^{t_4}(u)\cdots R_{2L-3,n}(u)\bar{R}_{2L-2,n}(u)\bar{R}_{2L-1,n}^{t_{2L-1}}(u)R_{2L,n}^{t_{2L}}(u),
\end{equation}
we find $|\Psi^D\rangle$ can be related to $|\Psi\rangle$ in (\ref{eq:4sitestate}) by the action of $t^D(-1)$:
\begin{equation}
|\Psi^D\rangle=t^D(-1)|\Psi\rangle.
\end{equation}
It can be easily shown that $t^D(u)$ commute with the two ABJM transfer matrices
\begin{equation}
[t^{D}(u),\tau(v)]=[t^{D}(u),\bar{\tau}(v)]=0.
\end{equation}
Therefore, $|\Psi^D\rangle$ also satisfy the chiral integrability condition:
\begin{equation}\label{eq:4sitedic}
\tau(u)|\Psi^D\rangle = \Pi \tau(u) \Pi |\Psi^D\rangle.
\end{equation}

\begin{figure}[h]
 \begin{center}
   \includegraphics[width=0.6\linewidth]{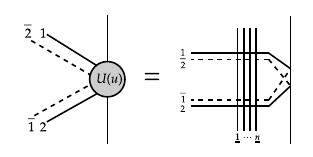}
 \end{center}
\caption{Fused $K$-matrix by dressing with monodromy $T$-matrices. \label{fig:dressedKbyT}}
\end{figure}

The generalization to more involved fused $K$-matrices dressed by monodromy $T$-matrices is straightforward. The resulting object can be viewed as a fused double-row monodromy, as shown in Fig.~\ref{fig:dressedKbyT}, with the explicit expression given by
\begin{equation}
U_{(\bar{1}2),(\underline{1}\cdots\underline{n})}(u)=R_{(\bar{1}2),(\underline{1}\cdots\underline{n})}(u)\tilde{K}_{(\bar{1}2)}(u)R_{(\underline{1}\cdots\underline{n}),(\bar{2}1)}(u),
\end{equation}
where the two fused $R$-matrices, which can also be seen as fused monodromy matrices, are given by
\begin{eqnarray}
&&R_{(\bar{1}2),(\underline{1}\cdots\underline{n})}(u)=\left[R_{2\underline{1}}(u)R_{2\underline{2}}(u)\cdots R_{2\underline{n}}(u)\right]
\left[R_{\bar{1}\underline{1}}(u)R_{\bar{1}\underline{2}}(u)\cdots R_{\bar{1}\underline{n}}(u)\right],\\
&&R_{(\underline{1}\cdots\underline{n}),(\bar{2}1)}(u)=\left[R_{\underline{n}\bar{2}}(u)\cdots R_{\underline{2}\bar{2}}(u) R_{\underline{1}\bar{2}}(u)\right]
\left[R_{\underline{n}1}(u)\cdots R_{\underline{2}1}(u)R_{\underline{1}1}(u)\right].
\end{eqnarray}
Then the four-site block is constructed from $U_{(\bar{1}2),(\underline{1}\cdots\underline{n})}(u)$ as
\begin{equation}
|\Phi^D(u)\rangle=\left[U(u)\right]^{a_1a_2}_{b_1b_2}|a_2,a_1,b_2,b_1\rangle\in{\rm{End}}\left(V_{\underline{1}}\otimes\cdots\otimes V_{\underline{n}}\right)\otimes V^{\otimes 4},
\end{equation}
from which the chiral integrable MPS is obtained.

\subsection{Multi-site generalization}
\begin{figure}[h]
 \begin{center}
   \includegraphics[width=0.4\linewidth]{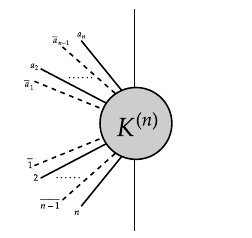}
 \end{center}
\caption{The $n$-fused $K^{(n)}$-matrix corresponding to the $n$-site chiral integrable MPS. \label{fig:nsite}}
\end{figure}

From the constructions of the two-site and four-site chiral integrable MPS described above, it is evident that a general $2n$-site chiral integrable MPS (bond-dimension-one) can be derived from the $n$-fused $K$-matrix: $K^{(n)}(u)=K_{(\bar{1}2\cdots\overline{n-1}n)}(u)$ shown in Fig.~\ref{fig:nsite}, where  $\{a_1 \cdots a_n\}$ represent an arbitrary permutation of $\{1 \cdots n\}$.

\begin{figure}[h]
 \begin{center}
   \includegraphics[width=0.8\linewidth]{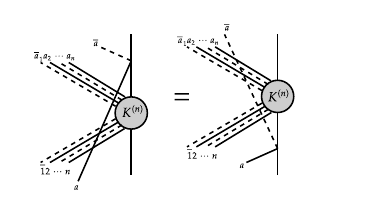}
 \end{center}
\caption{The $n$-fused boundary reflections. \label{fig:nfusesc}}
\end{figure}

The explicit form of the $n$-fused $K^{(n)}$-matrix depends on the specific scattering process, as long as the latter enables $K^{(n)}(u)$ satisfy the following fused RE, which is also graphically depicted in Fig.~\ref{fig:nfusesc},
\begin{equation}
\begin{split}
R_{(\bar{1}2\cdots\overline{n-1}n),a}(u-v)K^{(n)}(u)R_{a,(\bar{a}_1a_2\cdots\bar{a}_{n-1}a_n)}(u+v)K_{a}(v)\\
=K_a(v)R_{(\bar{1}2\cdots\overline{n-1}n),\bar{a}}(u+v)K^{(n)}(u)R_{\bar{a},(\bar{a}_1a_2\cdots \bar{a}_{n-1}a_n)}(u-v),
\end{split}
\end{equation}
where the fused $R$-matrices are
\begin{eqnarray}
&&R_{(\bar{1}2\cdots\overline{n-1}n),a}(u)=R_{na}(u)\bar{R}_{n-1,a}(u)\cdots R_{2a}(u)\bar{R}_{1a}(u),\\
&&R_{a,(\bar{a}_1a_2\cdots\bar{a}_{n-1}a_n)}(u)=\bar{R}_{a,a_1}(u)R_{a,a_2}(u)\cdots \bar{R}_{a,a_{n-1}}(u)R_{a,a_n}(u),
\end{eqnarray}
and similar for $R_{(\bar{1}2\cdots\overline{n-1}n),\bar{a}}(u)$ and $R_{\bar{a},(\bar{a}_1a_2\cdots \bar{a}_{n-1}a_n)}(u)$.

The $2n$-site block is then defined as
\begin{equation}
|\Phi^{(n)}(u)\rangle=\left[K^{(n)}(u)\right]^{i_1\cdots i_n}_{j_1\cdots j_n}|i_n\cdots i_1,j_n\cdots,j_1\rangle,
\end{equation}
from which we can construct the $2n$-site translationally invariant MPS,
\begin{equation}
|\Psi^{(n)}\rangle=|\Phi^{(n)}(-1)\rangle\otimes \cdots \otimes|\Phi^{(n)}(-1)\rangle.
\end{equation}
By construction, $|\Psi^{(n)}\rangle$ satisfy the chiral integrability condition.

\section{Overlap formula}\label{sec:overlap}

In this section, we propose exact formulas for the overlaps between the four-site chiral integrable MPS and Bethe states $\left \langle \Psi  | \mathbf{u}, \mathbf{w}, \mathbf{v} \right \rangle $. Overlap formulas between integrable boundary states and spin-chain eigenstates are subject to a set of selection rules, and the non-vanishing overlaps are proportional to the determinant of a Gaudin-like matrix. In the chiral integrable case considered here, one of the selection rules requires that the Bethe roots characterizing $\left | \mathbf{u}, \mathbf{w}, \mathbf{v} \right \rangle$ are parity symmetric, namely $\mathbf{u}=-\mathbf{u}$, $\mathbf{w}=-\mathbf{w}$, $\mathbf{v}=-\mathbf{v}$. We first present the norms of the Bethe states, the expression for the Gaudin determinant, and explain how these expressions simplify in the parity-symmetric case. We then introduce additional selection rules and present the overlap formulas $\left \langle \Psi  | \mathbf{u}, \mathbf{w}, \mathbf{v} \right \rangle $ for the symmetric $\tilde{K}^{S}_{a}$ and antisymmetric $\tilde{K}^{A}_{a}$ cases, respectively. The Bethe roots used to check chiral integrability, the selection rules, and the overlap formulas are collected in \autoref{roots}. 

\subsection{Norms of the Bethe states}

The norms of the Bethe states in the $SU(4)$-invariant alternating spin chain are given by
\begin{align}
\left \langle \mathbf{u}, \mathbf{w}, \mathbf{v} | \mathbf{u}, \mathbf{w}, \mathbf{v} \right \rangle
=&\left (\prod_{i<j}\frac{S\left ( u_i,u_j \right )}{S\left ( u_i^*,u_j^* \right )} \right )^{\frac{1}{2}}
\left (\prod_{i<j}\frac{S\left ( v_i,v_j \right )}{S\left ( v_i^*,v_j^* \right )} \right )^{\frac{1}{2}}
\left (\prod_{i<j}\frac{S\left ( w_i,w_j \right )}{S\left ( w_i^*,w_j^* \right )} \right )^{\frac{1}{2}}
\nonumber \\
&\times \left ( \prod_{j}\frac{1}{\partial_{u_j}p\left ( u_j \right ) }  \right ) 
\left ( \prod_{k}\frac{1}{\partial_{v_k}p\left ( v_k \right ) }  \right ) \det G, \label{norm}
\end{align}
where $\left \langle \mathbf{u}, \mathbf{w}, \mathbf{v} \right | \equiv \left | \mathbf{u}, \mathbf{w}, \mathbf{v} \right \rangle^{\dagger }$. 
Here $p(u)$ denotes the momentum of a magnon with rapidity $u$
\begin{align}
e^{i p(u)} = \frac{u+\frac{i}{2}}{u-\frac{i}{2}} \quad\Rightarrow\quad \partial_{u}p\left ( u \right ) =\frac{-1}{u^2+\frac{1}{4} }.\label{pupu}  
\end{align}
The determinant $\det G$ is the Gaudin determinant of the $SU(4)$ alternating spin chain, and its explicit form will be presented in the next subsection. 

\subsection{Gaudin determinant}\label{subsec:Gaudin}

The Gaudin matrix of the $SU(4)$ alternating spin chain is given by 
\begin{align}
G=
\begin{pmatrix}
\partial_{u_i}\phi_{u_j} & \partial_{u_i}\phi_{w_j} & \partial_{u_i}\phi_{v_j}\\
\partial_{w_i}\phi_{u_j} & \partial_{w_i}\phi_{w_j} & \partial_{w_i}\phi_{v_j}\\
\partial_{v_i}\phi_{u_j} & \partial_{v_i}\phi_{w_j} & \partial_{v_i}\phi_{v_j}
\end{pmatrix},\label{gaudinmatrix} 
\end{align}
where $\phi$s are defined through \eqref{bae}. 
\begin{align}
\partial_{u_i}\phi_{u_j}&=\delta_{ij}\left ({p}' \left ( u_i \right ) L
+\sum_{k=1}^{N_u} \varphi\left ( u_i, u_k \right )
+\sum_{k=1}^{N_w} \widetilde{\varphi}\left ( u_i, w_k \right ) \right )
-\varphi\left ( u_i, u_j \right ), \\
\partial_{u_i}\phi_{w_j}&=-\widetilde{\varphi}\left ( u_i, w_j \right ), \\
\partial_{u_i}\phi_{v_j}&=0, \\
\partial_{w_i}\phi_{u_j}&=-\widetilde{\varphi}\left ( w_i, u_j \right ), \\
\partial_{w_i}\phi_{w_j}&=\delta_{ij}\left (\sum_{k=1}^{N_w} \varphi\left ( w_i, w_k \right )
+\sum_{k=1}^{N_u} \widetilde{\varphi}\left ( w_i, u_k \right )
+\sum_{k=1}^{N_v} \widetilde{\varphi}\left ( w_i, v_k \right ) \right )
-\varphi\left ( w_i, w_j \right ), \\
\end{align}
where
\begin{align}
&\varphi\left ( u, v \right )=-i \frac{\partial }{\partial u} \log S\left ( u,v \right )
=\frac{2}{1+\left ( u-v \right )^2 } , \\
&\widetilde{\varphi}\left ( u, v \right )=-i \frac{\partial }{\partial u} \log \widetilde{S} \left ( u,v \right )
=\frac{-4}{1+4\left ( u-v \right )^2 }, \\
&\frac{\partial }{\partial u} \log S\left ( u,v \right )=-\frac{\partial }{\partial u} \log S\left ( v,u \right ), \\
&\frac{\partial }{\partial u} \log \widetilde{S}\left ( u,v \right )=-\frac{\partial }{\partial u} \log \widetilde{S}\left ( v,u \right ). 
\end{align}

For parity-symmetric states, the Gaudin determinant $\det G$ can be further factorized as $\det G = \det G_{+}  \det G_{-}$. We take parity-symmetric rapidities $\mathbf{u}=-\mathbf{u}$, $\mathbf{w}=-\mathbf{w}$, $\mathbf{v}=-\mathbf{v}$ ordered as:
\begin{align}
\mathbf{u} = \begin{cases}
\left \{ u_1, -u_1, u_2, -u_2, \cdots, u_{\frac{N_u}{2}}, -u_{\frac{N_u}{2}} \right \}   &  N_u \ \text{even} \\
\left \{ u_1, -u_1, u_2, -u_2, \cdots, u_{\left \lfloor \frac{N_u}{2} \right \rfloor}, -u_{\left \lfloor \frac{N_u}{2} \right \rfloor },0 \right \}   &  N_u \ \text{odd}
\end{cases} \label{root order}
\end{align}
and similarly for $\mathbf{w}$ and $\mathbf{v}$. 

As an illustration, we take $N_u, N_w, N_v$ to be odd integers. For even excitation numbers, the derivation proceeds analogously, with the corresponding rows and columns associated with the zero roots removed. For odd excitation numbers, each set of Bethe roots can be further divided into three parts, 
\begin{align}
&\mathbf{u} = \mathbf{u}^{+}\cup \mathbf{u}^{-}\cup \mathbf{u}^{0}, \\
&\mathbf{u}^{+}=\left \{ u_1, u_2, \cdots, u_{\left \lfloor \frac{N_u}{2}\right \rfloor } \right \}, \\
&\mathbf{u}^{-}=\left \{ -u_1, -u_2, \cdots, -u_{\left \lfloor \frac{N_u}{2}\right \rfloor } \right \}, \\
&\mathbf{u}^{0}=\left \{ 0 \right \}, 
\end{align}
and similarly for $\mathbf{w}$ and $\mathbf{v}$.
Then we can reorder the rows and columns of Gaudin matrix
\begin{align}
\det G&=\det \begin{pmatrix}
U^{+}_{u^{+}} &U^{+}_{w^{+}} &U^{+}_{v^{+}} &U^{+}_{u^{0}} &U^{+}_{w^{0}} &U^{+}_{v^{0}} &U^{+}_{u^{-}} &U^{+}_{w^{-}} &U^{+}_{v^{-}} \\
W^{+}_{u^{+}} &W^{+}_{w^{+}} &W^{+}_{v^{+}} &W^{+}_{u^{0}} &W^{+}_{w^{0}} &W^{+}_{v^{0}} &W^{+}_{u^{-}} &W^{+}_{w^{-}} &W^{+}_{v^{-}} \\
V^{+}_{u^{+}} &V^{+}_{w^{+}} &V^{+}_{v^{+}} &V^{+}_{u^{0}} &V^{+}_{w^{0}} &V^{+}_{v^{0}} &V^{+}_{u^{-}} &V^{+}_{w^{-}} &V^{+}_{v^{-}} \\
U^{0}_{u^{+}} &U^{0}_{w^{+}} &U^{0}_{v^{+}} &U^{0}_{u^{0}} &U^{0}_{w^{0}} &U^{0}_{v^{0}} &U^{0}_{u^{-}} &U^{0}_{w^{-}} &U^{0}_{v^{-}} \\
W^{0}_{u^{+}} &W^{0}_{w^{+}} &W^{0}_{v^{+}} &W^{0}_{u^{0}} &W^{0}_{w^{0}} &W^{0}_{v^{0}} &W^{0}_{u^{-}} &W^{0}_{w^{-}} &W^{0}_{v^{-}} \\
V^{0}_{u^{+}} &V^{0}_{w^{+}} &V^{0}_{v^{+}} &V^{0}_{u^{0}} &V^{0}_{w^{0}} &V^{0}_{v^{0}} &V^{0}_{u^{-}} &V^{0}_{w^{-}} &V^{0}_{v^{-}} \\
U^{-}_{u^{+}} &U^{-}_{w^{+}} &U^{-}_{v^{+}} &U^{-}_{u^{0}} &U^{-}_{w^{0}} &U^{-}_{v^{0}} &U^{-}_{u^{-}} &U^{-}_{w^{-}} &U^{-}_{v^{-}} \\
W^{-}_{u^{+}} &W^{-}_{w^{+}} &W^{-}_{v^{+}} &W^{-}_{u^{0}} &W^{-}_{w^{0}} &W^{-}_{v^{0}} &W^{-}_{u^{-}} &W^{-}_{w^{-}} &W^{-}_{v^{-}} \\
V^{-}_{u^{+}} &V^{-}_{w^{+}} &V^{-}_{v^{+}} &V^{-}_{u^{0}} &V^{-}_{w^{0}} &V^{-}_{v^{0}} &V^{-}_{u^{-}} &V^{-}_{w^{-}} &V^{-}_{v^{-}}
\end{pmatrix}, 
\end{align}
where $\left [ U^{+}_{w^{+}} \right ]_{ij}=\partial_{u^{+}_{i}}\phi_{w^{+}_{j}}$, and analogous definitions apply to the other blocks. Then we can add the rows and subtract the columns to factorize the determinant.
\begin{align}
\resizebox{\textwidth}{!}{$
\det G=\det
\begin{pmatrix}
\color{red}{U^{+}_{u^{+}} + U^{-}_{u^{+}}} & \color{red}{U^{+}_{w^{+}} + U^{-}_{w^{+}}} & \color{red}{U^{+}_{v^{+}} + U^{-}_{v^{+}}} 
& \color{red}{2 U^{+}_{u^{0}}} & \color{red}{2U^{+}_{w^{0}}} & \color{red}{2U^{+}_{v^{0}}} 
& 0 & 0 & 0 \\
\color{red}{W^{+}_{u^{+}} + W^{-}_{u^{+}}} & \color{red}{W^{+}_{w^{+}} + W^{-}_{w^{+}}} & \color{red}{W^{+}_{v^{+}} + W^{-}_{v^{+}}} 
& \color{red}{2W^{+}_{u^{0}} } & \color{red}{2W^{+}_{w^{0}}} & \color{red}{2W^{+}_{v^{0}}} 
& 0 & 0 & 0 \\
\color{red}{V^{+}_{u^{+}} + V^{-}_{u^{+}}} & \color{red}{V^{+}_{w^{+}} + V^{-}_{w^{+}}} & \color{red}{V^{+}_{v^{+}} + V^{-}_{v^{+}}} 
& \color{red}{2V^{+}_{u^{0}}} & \color{red}{2V^{+}_{w^{0}}} & \color{red}{2V^{+}_{v^{0}}} 
& 0 & 0 & 0 \\
\color{red}{U^{0}_{u^{+}}} & \color{red}{U^{0}_{w^{+}}} & \color{red}{U^{0}_{v^{+}}} & \color{red}{U^{0}_{u^{0}}} & \color{red}{U^{0}_{w^{0}}} & \color{red}{U^{0}_{v^{0}}} 
& 0 & 0 & 0 \\
\color{red}{W^{0}_{u^{+}}} & \color{red}{W^{0}_{w^{+}}} & \color{red}{W^{0}_{v^{+}}} & \color{red}{W^{0}_{u^{0}}} & \color{red}{W^{0}_{w^{0}}} & \color{red}{W^{0}_{v^{0}}} 
& 0 & 0 & 0 \\
\color{red}{V^{0}_{u^{+}}} & \color{red}{V^{0}_{w^{+}}} & \color{red}{V^{0}_{v^{+}}} & \color{red}{V^{0}_{u^{0}}} & \color{red}{V^{0}_{w^{0}}} & \color{red}{V^{0}_{v^{0}}} 
& 0 & 0 & 0 \\
U^{-}_{u^{+}} & U^{-}_{w^{+}} & U^{-}_{v^{+}} & U^{-}_{u^{0}} & U^{-}_{w^{0}} & U^{-}_{v^{0}} 
& \color{blue}{U^{-}_{u^{-}} - U^{-}_{u^{+}}} & \color{blue}{U^{-}_{w^{-}} - U^{-}_{w^{+}}} & \color{blue}{U^{-}_{v^{-}} - U^{-}_{v^{+}}} \\
W^{-}_{u^{+}} & W^{-}_{w^{+}} & W^{-}_{v^{+}} & W^{-}_{u^{0}} & W^{-}_{w^{0}} & W^{-}_{v^{0}} 
& \color{blue}{W^{-}_{u^{-}} - W^{-}_{u^{+}}} & \color{blue}{W^{-}_{w^{-}} - W^{-}_{w^{+}}} & \color{blue}{W^{-}_{v^{-}} - W^{-}_{v^{+}}} \\
V^{-}_{u^{+}} & V^{-}_{w^{+}} & V^{-}_{v^{+}} & V^{-}_{u^{0}} & V^{-}_{w^{0}} & V^{-}_{v^{0}} 
& \color{blue}{V^{-}_{u^{-}} - V^{-}_{u^{+}}} & \color{blue}{V^{-}_{w^{-}} - V^{-}_{w^{+}}} & \color{blue}{V^{-}_{v^{-}} - V^{-}_{v^{+}}}
\end{pmatrix}$}
\end{align}
in which the red block is $G_{+}$, blue block is $G_{-}$.

\subsection{Exact overlap formulas}

As constructed above, we have chiral integrable boundary states of the form of MPS with bond dimension one
\begin{align}
&\left | \Psi  \right \rangle = \left | \Phi \left ( -1 \right )  \right \rangle^{\otimes \frac{L}{2}}, \\
&\left | \Phi \left ( -1 \right )  \right \rangle 
= \left [ \tilde{K}_{\left ( \bar{1}2 \right ) }\left ( -1 \right )  \right ]^{i_1,i_2}_{j_1,j_2}
\left | i_1,i_2,j_1,j_2, \right \rangle, \\
&\tilde{K}_{\left ( \bar{1}2 \right ) }\left ( -1 \right )
=\tilde{K}_{2}R_{\bar{1}\bar{2}}\left ( -1 \right )\tilde{K}_{\bar{1} }
=\tilde{K}_{2}R_{\bar{1}\bar{2}}\left ( -1 \right )\left ( \tilde{K}_{1} \right )^{-1},
\end{align}
where $\tilde{K}_{a}$ is a symmetric or anti-symmetric constant solution of the SNP-type RE.

Note that here we take the Bethe roots to be ordered as in $\eqref{root order}$.

For the symmetric $\tilde{K}^{S}_{a}$, an additional selection rule emerges for the overlap to be non-vanishing, beyond the chiral pairing structure of the Bethe roots. Specifically, the excitation numbers $N_u, N_w, N_v$ must all be even integers. The exact overlap formula then takes the form
\begin{align}
\left \langle \Psi^{S} | \mathbf{u}, \mathbf{w}, \mathbf{v} \right \rangle
=&\left ( -1 \right )^{\left ( \frac{N_u+N_v}{2}+N_w \right ) }\left ( -2 \right )^{\frac{L}{2}}
\prod_{i}^{\frac{N_u}{2}}u_i\left ( u_i-\frac{i}{2}\right )
\prod_{i}^{\frac{N_v}{2}}v_i\left ( v_i-\frac{i}{2}\right )
\prod_{i}^{\frac{N_w}{2}}\frac{w_i}{w_i+\frac{i}{2}}\nonumber \\
&\times \left [ \left ( P_u P_w P_v \right )^{-\frac{L}{2} }
P_u^{\frac{N_u}{2} }P_w^{\frac{N_w}{2} }P_v^{\frac{N_v}{2} } \right ]^{*}
\times \det G_{+},
\end{align}
where
\begin{align}
P_u=S_{\left ( 4 \right ) }S_{\left ( 3 \right ) }^{-2}S_{\left ( 2 \right ) },\  
P_w=S_{\left ( 3 \right ) }S_{\left ( 2 \right ) }^{-2}S_{\left ( 1 \right ) },\ 
P_v=S_{\left ( 2 \right ) }S_{\left ( 1 \right ) }^{-2},
\end{align}
in which $S_{\left ( i \right )}$ denotes the $i$-th bottom-right principal minor of $\tilde{K}^{S}_{a}$, explicitly given by
\begin{align}
&S_{\left ( 4 \right ) }=\det \tilde{K}^{S}_{a}
=\det \begin{pmatrix}
s_{11}  &s_{12}  &s_{13}  &s_{14} \\
s_{12}  &s_{22}  &s_{23}  &s_{24} \\
s_{13}  &s_{23}  &s_{33}  &s_{34} \\
s_{14}  &s_{24}  &s_{34}  &s_{44}
\end{pmatrix},\\
&S_{\left ( 3 \right ) }=\det \begin{pmatrix}
s_{22}  &s_{23}  &s_{24} \\
s_{23}  &s_{33}  &s_{34} \\
s_{24}  &s_{34}  &s_{44}
\end{pmatrix},\\
&S_{\left ( 2 \right ) }=\det \begin{pmatrix}
s_{33}  &s_{34} \\
s_{34}  &s_{44}
\end{pmatrix},\\
&S_{\left ( 1 \right ) }=\det \begin{pmatrix}
s_{44}
\end{pmatrix}=s_{44}.
\end{align}
Thus the normalized overlap formula is given by
\begin{align}
\frac{\left \langle \Psi^{S} | \mathbf{u}, \mathbf{w}, \mathbf{v} \right \rangle}{\sqrt{\left \langle \mathbf{u}, \mathbf{w}, \mathbf{v} | \mathbf{u}, \mathbf{w}, \mathbf{v} \right \rangle} }
=&\left ( -1 \right )^{\left ( \frac{N_u+N_v}{2}+N_w \right ) }\left ( -2 \right )^{\frac{L}{2}}
\times \prod_{i=1}^{\frac{N_u}{2}} \left | \frac{u_i+\frac{i}{2}}{u_i-\frac{i}{2}} \right |^{\frac{1}{2}}
\prod_{i=1}^{\frac{N_v}{2}} \left | \frac{v_i+\frac{i}{2}}{v_i-\frac{i}{2}} \right |^{\frac{1}{2}}
\prod_{i=1}^{\frac{N_w}{2}} \left | \frac{w_i+\frac{i}{2}}{w_i-\frac{i}{2}} \right |^{\frac{1}{2}} \nonumber \\
&\times \prod_{i=1}^{\frac{N_u}{2}}\frac{u_i}{u_i+\frac{i}{2}}
\prod_{i=1}^{\frac{N_v}{2}}\frac{v_i}{v_i+\frac{i}{2}}
\prod_{i=1}^{\frac{N_w}{2}}\frac{w_i}{w_i+\frac{i}{2}}\nonumber \\
&\times \left [ \left ( P_u P_w P_v \right )^{-\frac{L}{2} }
P_u^{\frac{N_u}{2} }P_w^{\frac{N_w}{2} }P_v^{\frac{N_v}{2} } \right ]^{*}
\times \sqrt{\frac{\det G_{+}}{\det G_{-}}}.
\end{align}

For the antisymmetric case $\tilde{K}^{A}_{a}$, we find a different additional selection rule compared to the symmetric case, which shows that  $L,N_u,N_w, N_v$ must satisfy 
\begin{align}
L+N_w=N_u+N_v. 
\end{align}
The corresponding exact overlap formula is given by
\begin{align}
\left \langle \Psi^{A} | \mathbf{u}, \mathbf{w}, \mathbf{v} \right \rangle
=&\left ( -2 \right )^{\frac{L}{2}+N_w-2\left \lceil \frac{N_u}{2}\right \rceil -2 \left \lceil \frac{N_v}{2}\right \rceil} \prod_{i = 1}^{\left \lfloor \frac{N_u}{2}  \right \rfloor }\frac{u_i-\frac{i}{2}}{u_i}
\prod_{i = 1}^{\left \lfloor \frac{N_v}{2}  \right \rfloor }\frac{v_i-\frac{i}{2}}{v_i}
\prod_{i = 1}^{\left \lfloor \frac{N_w}{2}  \right \rfloor }w_i\left ( w_i-\frac{i}{2}\right )\nonumber \\
&\times \left [ \left ( F_u F_w F_v \right )^{-\frac{L}{2} }
F_u^{\frac{N_u}{2} }F_w^{\frac{N_w}{2} }F_v^{\frac{N_v}{2} } \right ]^{*}
\times \det G_{+},
\end{align}
where
\begin{align}
F_u=A_{\left ( 4 \right ) }A_{\left ( 2 \right ) },\ F_w=A_{\left ( 2 \right ) }^{-2},\ F_v=A_{\left ( 2 \right ) },
\end{align}
in which $A_{\left ( i \right )}$ is the $i$-th bottom-right principal minor of $\tilde{K}^{A}_{a}$
\begin{align}
&A_{\left ( 4 \right ) }=\det \tilde{K}^{A}_{a}=\mathrm{Pf} \left ( \tilde{K}^{A}_{a} \right )^{2},\\
&A_{\left ( 2 \right ) }=\det \begin{pmatrix}
0  & a_{34}\\
-a_{34}  & 0
\end{pmatrix},\\
&A_{\left ( 3 \right ) }=A_{\left ( 1 \right ) }=0.
\end{align}
This yields the normalized overlap formula
\begin{align}
\frac{\left \langle \Psi^{A} | \mathbf{u}, \mathbf{w}, \mathbf{v} \right \rangle}{\sqrt{\left \langle \mathbf{u}, \mathbf{w}, \mathbf{v} | \mathbf{u}, \mathbf{w}, \mathbf{v} \right \rangle} }
=&\left ( -2 \right )^{\frac{L}{2}+N_w-2\left \lceil \frac{N_u}{2}\right \rceil -2 \left \lceil \frac{N_v}{2}\right \rceil} \times2^{N_{u}-2\left \lfloor \frac{N_u}{2}\right \rfloor +N_{v}-2\left \lfloor \frac{N_v}{2}\right \rfloor} \nonumber \\
&\times
\prod_{i=1}^{\left \lfloor \frac{N_u}{2} \right \rfloor } \left | \frac{u_i+\frac{i}{2}}{u_i-\frac{i}{2}} \right |^{\frac{1}{2}}
\prod_{i=1}^{\left \lfloor \frac{N_v}{2} \right \rfloor } \left | \frac{v_i+\frac{i}{2}}{v_i-\frac{i}{2}} \right |^{\frac{1}{2}}
\prod_{i=1}^{\left \lfloor \frac{N_w}{2} \right \rfloor } \left | \frac{w_i+\frac{i}{2}}{w_i-\frac{i}{2}} \right |^{\frac{1}{2}}\nonumber \\
&\times
\prod_{i = 1}^{\left \lfloor \frac{N_u}{2}  \right \rfloor }\frac{1}{u_i\left ( u_i+\frac{i}{2} \right ) }
\prod_{i = 1}^{\left \lfloor \frac{N_v}{2}  \right \rfloor }\frac{1}{v_i\left ( v_i+\frac{i}{2} \right ) }
\prod_{i = 1}^{\left \lfloor \frac{N_w}{2}  \right \rfloor }w_i\left ( w_i-\frac{i}{2}\right )\nonumber \\
&\times \left [ \left ( F_u F_w F_v \right )^{-\frac{L}{2} }
F_u^{\frac{N_u}{2} }F_w^{\frac{N_w}{2} }F_v^{\frac{N_v}{2} } \right ]^{*}
\times \sqrt{\frac{\det G_{+}}{\det G_{-}}} .
\end{align}

Furthermore, by introducing an inner space carrying the $SU(4)$ fundamental representation through the $R$-matrix into the c-number $K$-matrices, we obtain chiral integrable MPS of bond dimension four
\begin{align}
\left | \Psi ^{D} \right \rangle = t^{D}\left ( -1 \right ) \left | \Psi  \right \rangle.
\end{align}
This transfer matrix evaluated at $u=-1$ coincides exactly with the transfer matrix of the ABJM spin chain at the same spectral parameter
\begin{align}
t^{D}\left ( -1 \right ) =\tau \left ( -1 \right ).
\end{align}
As a result, the overlap $\left \langle \Psi^{D} | \mathbf{u}, \mathbf{w}, \mathbf{v} \right \rangle$ can be readily obtained as
\begin{align}
\left \langle \Psi^{D} | \mathbf{u}, \mathbf{w}, \mathbf{v} \right \rangle
=\left \langle \Psi |\tau\left ( -1 \right ) | \mathbf{u}, \mathbf{w}, \mathbf{v} \right \rangle
=\Lambda \left ( -1 \right ) \left \langle \Psi | \mathbf{u}, \mathbf{w}, \mathbf{v} \right \rangle,
\end{align}
where $\tau\left ( -1 \right ) \left | \Psi  \right \rangle =\tau^{\dagger }\left ( -1 \right ) \left | \Psi  \right \rangle$ has been used. 

\section{Chiral integrable subspaces for small $L$}\label{sec:subspace}

To achieve a better understanding of the complete classification and construction of chiral integrable boundary states, we performed some explicit calculations for $L=2,3$. 
$\left | \mathcal{B} \right \rangle $ is chiral integrable if it satisfies $\eqref{untwisted condition}$. 
Then expanding the transfer matrices in $u$
\begin{align}
\tau(u)&=\sum_{s=0}^{2L} u^{s}\hat{\mathcal{O}}_s,\\
\bar{\tau}(-u-2)&=\sum_{s=0}^{2L} u^{s}\hat{\overline{\mathcal{O}}}_s,
\end{align}
gives an equivalent condition for $\left | \mathcal{B} \right \rangle $ to be chiral integrable
\begin{align}
\hat{\mathcal{O} }_s\left | \mathcal{B}\right \rangle  = \hat{\overline{\mathcal{O}}}_s \left | \mathcal{B}\right \rangle, s & = 0, \cdots, 2L.\label{ohs=ohbs}
\end{align}
For small $L$, we can write out these operators $\hat{\mathcal{O} }_s, \hat{\overline{\mathcal{O}}}_s$ in terms of permutation operators and trace operators explicitly. Below, we solve these conditions and present the results separately for $L=2$ and $L=3$.

For $L=2$, conditions labeled by $s=2,3,4$ are automatically satisfied. Two non-trivial constraints are imposed for $\left | \mathcal{B} \right \rangle$ to be chiral integrable:
\begin{align}
s&=1\quad
\left (\mathbb{K}_{1\bar{1}}\mathbb{K}_{1\bar{2}}+\mathbb{K}_{2\bar{2}}\mathbb{K}_{\bar{1}2}-\mathbb{K}_{1\bar{1}}\mathbb{K}_{\bar{1}2}-\mathbb{K}_{2\bar{2}}\mathbb{K}_{1\bar{2}}\right)
\left | \mathcal{B} \right \rangle \nonumber \\
&\quad\quad\quad=\left (\mathbb{K}_{1\bar{2}}\mathbb{K}_{1\bar{1}}+\mathbb{K}_{\bar{1}2}\mathbb{K}_{2\bar{2}}-\mathbb{K}_{\bar{1}2}\mathbb{K}_{1\bar{1}}-\mathbb{K}_{1\bar{2}}\mathbb{K}_{2\bar{2}}\right)
\left | \mathcal{B} \right \rangle, \label{s=1}   \\
s&=0\quad
\left (\mathbb{K}_{1\bar{1}}\mathbb{P}_{12}\mathbb{K}_{1\bar{2}}-2 \mathbb{P}_{12}\mathbb{K}_{\bar{1}2}-2 \mathbb{P}_{12}\mathbb{K}_{1\bar{2}}\right)
\left | \mathcal{B} \right \rangle \nonumber \\
&\quad\quad\quad=\left (\mathbb{K}_{1\bar{2}}\mathbb{P}_{\bar{1}\bar{2}}\mathbb{K}_{2\bar{2}}-2 \mathbb{K}_{\bar{1}2}\mathbb{K}_{1\bar{1}}-2 \mathbb{K}_{1\bar{2}}\mathbb{K}_{2\bar{2}}\right)
\left | \mathcal{B} \right \rangle. \label{s=0}
\end{align}
Now the full Hilbert space has dimension 256. We first consider
\begin{align}
\left | \mathcal{B}  \right \rangle =\sum_{A,B,C,D=1}^{4} c_{A,B,C,D} \left | A,\bar{B} ,C,\bar{D} \right \rangle.
\end{align}
Solving $\eqref{s=1}$ and $\eqref{s=0}$, we find that among the 256 coefficients only 196 are independent \footnote{Actually here $\eqref{s=0}$ is sufficient to get the 196-dimensional subspace.}. Consequently, the space of chiral integrable boundary states forms a 196-dimensional subspace of the full Hilbert space.

We next consider a two-site ansatz of the form
\begin{align}
    \left | \mathcal{B}_{M} \right \rangle =\left ( M^{i}_{j}\left | i,j \right \rangle  \right )^{\otimes L}
\end{align}
where $M \in \mathbb{C}^{4 \times 4}$ is a constant matrix with 16 unknown parameters. For any such two-site boundary state, the condition $\eqref{s=1}$ is automatically satisfied. Substituting this ansatz into $\eqref{s=0}$, we solve for the admissible matrices $M$.

We find that all resulting matrices $M$ satisfy one of the four types of RE in ABJM spin chain, which is a SP-type one, 
\begin{align}
R_{12}\left ( u-v \right ) M_{1}\left ( u \right )R_{12}\left ( u+v \right ) M_{2}\left ( v \right )
=M_{2}\left ( v \right )R_{12}\left ( u+v \right )M_{1}\left ( u \right ) R_{12}\left ( u-v \right ),\label{re}
\end{align}
except for the solution proportional to the identity matrix, which satisfies all four types of RE. Here, the four types of RE are classified according to the different combinations of $R$ and $\bar{R}$ appearing in the equation. However, the converse statement does not hold: not every solution of the RE $\eqref{re}$ can generate a two-site chiral integrable boundary state. These can be seen from the following analysis.

By writing out the indices in $\eqref{s=0}$, we obtain an equation for the matrix elements $M^{i}_{j}$, 
\begin{align}
&\left [\mathbb{K}_{1\bar{1}}\mathbb{P}_{12}\mathbb{K}_{1\bar{2}}-2 \mathbb{P}_{12}\mathbb{K}_{\bar{1}2}-2 \mathbb{P}_{12}\mathbb{K}_{1\bar{2}}\right]^{k_{1},l_{1},k_{2},l_{2}}_{i_{1},j_{1},i_{2},j_{2}}
M^{i_1}_{j_1}M^{i_2}_{j_2}\left | k_{1},l_{1},k_{2},l_{2} \right \rangle \nonumber \\
&\quad = \left [\mathbb{K}_{1\bar{2}}\mathbb{P}_{\bar{1}\bar{2}}\mathbb{K}_{2\bar{2}}-2 \mathbb{K}_{\bar{1}2}\mathbb{K}_{1\bar{1}}-2 \mathbb{K}_{1\bar{2}}\mathbb{K}_{2\bar{2}}\right]^{k_{1},l_{1},k_{2},l_{2}}_{i_{1},j_{1},i_{2},j_{2}}
M^{i_1}_{j_1}M^{i_2}_{j_2}\left | k_{1},l_{1},k_{2},l_{2} \right \rangle.
\end{align}
Then by removing the basis vector and using the component forms $\mathbb{I}_{a_1a_2}^{b_1b_2}=\delta_{a_1}^{b_1}\delta_{a_2}^{b_2}$, $
\mathbb{P}_{a_1a_2}^{b_1b_2}=\delta_{a_1}^{b_2}\delta_{a_2}^{b_1}$, $\mathbb{K}_{a_1\bar{a}_2}^{b_1\bar{b}_2}=\delta^{{b}_1\bar{b}_2}\delta_{a_1\bar{a}_2}$ in (\ref{eq:components}), we obtain
\begin{align}
&\mathrm{tr}\left ( M^2 \right ) \delta^{k_1,l_1}\delta^{k_2,l_2}
-2 \left [ M^2 \right ]^{k_1}_{l_1}\delta^{k_2,l_2}
-2\left [ M^2 \right ]^{k_2}_{l_2}\delta^{k_1,l_1} \nonumber \\
&\quad = \left ( \mathrm{tr}M \right )^{2}\delta^{k_1,l_2}\delta^{k_2,l_1}
-2\left ( \mathrm{tr}M \right )\left [ M \right ]^{k_2}_{l_1}\delta^{k_1,l_2}
-2\left ( \mathrm{tr}M \right )\left [ M \right ]^{k_1}_{l_2}\delta^{k_2,l_1}.\label{s0index}
\end{align}
Similarly, by expanding the indices in $\eqref{re}$ and simplifying the expression, we arrive 
\begin{align}
\left [ M^{2} \right ]^{k_1}_{l_2}\delta^{k_2}_{l_1} & = \left [ M^{2} \right ]^{k_2}_{l_1}\delta^{k_1}_{l_2}.\label{indexre}
\end{align}
Below, we present the solutions to $\eqref{s0index}$ and show that they automatically satisfy $\eqref{indexre}$. 

Starting from $\eqref{s0index}$, setting $l_{1}=k_{1}$ and summing over this index gives
\begin{align}
\left [ M^{2} \right ]^{k_2}_{l_2} & = \frac{\mathrm{tr}M}{2} \left [ M \right ]^{k_2}_{l_2}
+\frac{2 \mathrm{tr}\left ( M^{2} \right )-\left ( \mathrm{tr}M \right )^{2}}{8} \delta^{k_2,l_2}. \label{3}
\end{align}
If $\mathrm{tr}M=0$, we obtain
\begin{align}
\left [ M^{2} \right ]^{k}_{l} = \frac{\mathrm{tr}\left ( M^{2} \right )}{4}\delta^{k,l},
\end{align}
which clearly satisfies $\eqref{indexre}$.
If $\mathrm{tr}M\ne 0$, substituting $\eqref{3}$ into $\eqref{s0index}$ yields
\begin{align}
\frac{1}{2} \left ( \mathrm{tr} M \right ) \delta^{k_1,l_1}\delta^{k_2,l_2}
-\left [ M \right ]^{k_{1}}_{l_{1}}\delta^{k_2,l_2}
-\left [ M \right ]^{k_{2}}_{l_{2}}\delta^{k_1,l_1}\nonumber \\
=\left ( \mathrm{tr} M \right ) \delta^{k_1,l_2}\delta^{k_2,l_1}
-2 \left [ M \right ]^{k_{2}}_{l_{1}}\delta^{k_1,l_2}
-2 \left [ M \right ]^{k_{1}}_{l_{2}}\delta^{k_2,l_1}.
\end{align}
Set $k_1=l_1=k_2=l_2=i$, we get
\begin{align}
\left [ M \right ]^{i}_{i}=\frac{\mathrm{tr}M}{4},i=1,2,3,4.
\end{align}
Setting $k_1=k_2=i,l_1=l_2=j,i\ne j$ gives
\begin{align}
\left [ M \right ]^{i}_{j}=0.
\end{align}
Thus
\begin{align}
\left [ M \right ]^{k}_{l}&=\frac{\mathrm{tr}M}{4} \delta^{k,l},\\
\left [ M^2 \right ]^{k}_{l}&=\frac{\mathrm{tr}\left ( M^2 \right )}{4} \delta^{k,l}.
\end{align}
which again clearly satisfies $\eqref{indexre}$. In particular, the constant solutions to the SP-type RE $\eqref{re}$ \cite{Arnaudon:2004sd} are more than the set of matrices $M$ obtained above.

Furthermore, we investigate whether the chiral integrable boundary states we have constructed so far can span the full 196-dimensional chiral integrable subspace. To this end, we randomly sample the free parameters appearing in the constructed states and generate a sufficiently large set of vectors. By examining their linear independence, we determine the dimension of the subspace they span.

Combining the chiral integrable basis states found in \cite{Liu:2025uiu}
\begin{align}
\left | \mathcal{B} \right \rangle = \begin{cases}
\left |A\bar{B}\cdots A\bar{B} \right \rangle, A\ne B  & \text{Any}\ L \\
\left |A\bar{B} A\bar{D}\cdots A\bar{B} A\bar{D} \right \rangle, A, B, D\ \text{are\ all\ mutually\ distinct}  & L\ \text{even}\\
\left |A\bar{B} C\bar{B}\cdots A\bar{B} C\bar{B} \right \rangle, A, B, C\ \text{are\ all\ mutually\ distinct}  & L\ \text{even}\\
\left |A\bar{B} C\bar{D}\cdots A\bar{B} C\bar{D} \right \rangle, A, B, C, D\ \text{are\ all\ mutually\ distinct}  & L\ \text{even}\end{cases}
\label{4types}
\end{align}
where $A,C\in \{1,2,3,4\}$,$\bar{B},\bar{D}\in \{\bar{1},\bar{2},\bar{3},\bar{4}\}$,
MPS with bond dimension one $\left | \Psi  \right \rangle$, higher-bond-dimension MPS $\left | \Psi^{D}  \right \rangle $, and the two-site chiral integrable boundary states $\left | \mathcal{B}_{M} \right \rangle$ obtained above, we find that these states span a 142-dimensional subspace. Thus there are still new constructions to be found. 

A more refined analysis reveals that, although the subspaces generated by bond-dimension-one MPS $\left | \Psi  \right \rangle$ and bond-dimension-four MPS $t^{D}(-1)\left | \Psi  \right \rangle$ depend mildly on whether the $K$-matrix is symmetric or antisymmetric, their union does not enlarge the resulting chiral integrable subspace. Specifically, for antisymmetric $K$-matrices, each class of MPS spans a 21-dimensional subspace, and their union has dimension 22. For symmetric $K$-matrices, each class spans an 85-dimensional subspace, and their union has dimension 86. Nevertheless, the subspaces generated separately by bond-dimension-one MPS, bond-dimension-four MPS, and the two-site chiral integrable boundary states $\left | \mathcal{B}_{M} \right \rangle$ are identical and have dimension 106.

We now turn to the case $L = 3$. In this case, the conditions labeled by $s = 6,5,4$ are automatically satisfied for all states. Solving the remaining constraints, we find that the resulting chiral integrable subspace has dimension 616. Furthermore, the condition $s = 3$ is satisfied for any two-site boundary state. Solving the remaining conditions $s = 2, 1, 0$, we find that all admissible matrices M are
$\mathrm{rank}(M) = 1$ ,
and that each of them obeys $\eqref{re}$. Conversely, only rank-one solutions of $\eqref{re}$ are capable of generating chiral integrable boundary states. 

An interesting observation is that although $\eqref{eq:mixre}$ admits no non-trivial solutions for generic values of $u$ and $v$, non-invertible solutions do exist when $u=-1$. In this case, the corresponding matrices $K_1(-1)$ are all of rank one. Moreover, numerical checks show that the two-site states generated from these $K_1(-1)$ are chiral integrable for system sizes $L=2$ and $L=3$. This indicates that, although the invertibility condition on $\tilde{K}_a$ required in the construction of two-site chiral integrable MPS in Subsection~\ref{subsec:two-site} is not satisfied, the construction remains formally valid, at least for small $L$. Also note that as mentioned above, numerical analysis shows that for $L=2$ these two-site integrable MPS span the same subspace as the four-site integrable MPS.

\section{Conclusion} \label{sec:conclusion}
In this paper, we present a general construction of chiral integrable MPS in ABJM spin chain based on boundary integrability, all of which correspond to $K$-matrix solutions of specific RE. Concretely, for the two-site case, we focus on two fundamental boundary reflections: the SP-type associated with the untwisted RE, and the SNP-type associated with the twisted one. By analyzing a mixed SP-SNP RE, we obtain a formal construction of two-site chiral integrable product states. However, finding non-trivial analytic solutions to these mixed RE is difficult. We therefore directly analyze the chiral integrability condition for two-site states. From this approach, we identify integrable states for small site numbers ($L=2$ and $L=3$). Remarkably, all these states correspond to a subset of solutions of the SP-type RE. Moreover, we find that, the product states arising from solutions of the mixed SP-SNP RE at special values of the spectral parameter exhibit chiral integrability. For the four-site and general $2n$-site cases, we employ the fusion procedure, which enables us to construct chiral integrable states from fused $K$-matrices. These fused $K$-matrices are solutions to specific fused RE, which correspond to certain carefully designed scattering processes composed of the fundamental SP- and SNP-type boundary reflections. We further extend the construction from chiral integrable product states to chiral integrable MPS with bond dimension greater than one. These chiral integrable MPS originate from operator-valued $K$-matrices that carry internal degrees of freedom. Such $K$-matrices can be obtained by dressing an ordinary c-number $K$-matrix with an $R$-matrix or more generally a double-row monodromy matrix.

For the obtained four-site chiral integrable states, we further investigate their overlaps with Bethe eigenstates. The construction of integrable four-site states relies on the solutions of the SNP-type RE. We demonstrate that the allowed invertible solutions fall into two classes: symmetric and antisymmetric constant matrices, which correspond to two distinct types of four-site chiral integrable MPS. Consistent with previous studies, both types of four-site integrable MPS yield exact overlap formulas, namely, square root of a ratio of two Gaudin-like determinants multiplied by scalar factors including ones originating from the $K$-matrices (specifically, their determinants and minors). We also study the chiral integrable subspace numerically by collecting the chiral integrable states derived in this work together with those from \cite{Liu:2025uiu}, and examine the dimension of this subspace and its variation under the action of the transfer matrices. For $L=2$, all currently known chiral integrable boundary states span a $142$-dimensional subspace, whereas the full chiral integrable subspace has dimension $196$.

Based on the present work, several directions are worth pursuing for future research. First, it would be highly desirable to provide a general and rigorous proof for our overlap formula of four-site integrable product states, as well as its further generalization to $2n$-site integrable chiral MPS. Second, a natural counterpart to the chiral construction here is to use the similar fusion method to construct general $2n$-site achiral integrable MPS and investigate their overlaps. Third, our numerical analysis for $L=2$ indicates that the currently known chiral integrable boundary states do not exhaust the full chiral integrable subspace, suggesting that additional, yet unidentified, constructions may exist. This motivates a further systematic exploration of possible new classes of chiral integrable boundary states. Finally, our current analysis of chiral integrable subspaces is limited to small site numbers, and thus an analytic treatment for general $L$ is obviously of great interest. It would be interesting to examine whether the construction of complete basis of integrable boundary states \cite{Ekhammar:2023iph}, known for spin chains with boundary twist, admits an analogue in the untwisted case.

\section*{Acknowledgments}

NB, MS and JW thank the participants of the external program [{\textbf{APCTP-2025-E12}}], {\textit {Challenges in integrabililty}}, held at Wuhan, China for warm discussions. We thank the Innovation Academy for Precision Measurement Science and Technology for warm hospitality during this event.
YL thank Yu-Xuan Zhang for very helpful discussions. 
This work is supported  by the National Natural Science Foundation of China (NSFC) Grants No.~12375006, 12165002, 11975164,  12247103, and 
Tianjin University Self-Innovation Fund Extreme Basic Research Project Grant No.~2025XJ21-0007.
Some numerical calculations were conducted on the CJQS-HPC platform at Tianjin University.

\appendix

\section{Bethe roots}
\label{roots}

Besides the roots found in \cite{Yang:2021hrl,Jiang:2023cdm,Liu:2025uiu}, we solve the Bethe ansatz equations (BAEs) of the $SU(4)$ alternating spin chain to obtain additional solutions, which are used to numerically test our results. The roots employed in our analysis are listed in \autoref{paired roots}, and solutions obtained merely by swapping the values of $N_u$ and $N_v$ are omitted. Note that, for the excitation numbers of interest, it is computationally more efficient to impose the pairing structure of the roots, $\mathbf{u}=-\mathbf{u}, \mathbf{v}=-\mathbf{v}, \mathbf{w}=-\mathbf{w}$, and then solve the BAEs directly, rather than solving the corresponding Q-system. We have also carefully excluded unphysical solutions. 

\begin{longtable}{>{\centering\arraybackslash}p{2.5cm}|p{11.5cm}}
\caption{Parity symmetric roots}
\label{paired roots}\\

\hline
$(L,N_u,N_v,N_w)$ & [$\mathbf{u},\mathbf{v},\mathbf{w}$] \\
\hline
\hline
\endfirsthead

\hline
$(L,N_u,N_v,N_w)$ & [$\mathbf{u},\mathbf{v},\mathbf{w}$] \\
\hline
\hline
\endhead


$(2,1,0,0)$ & $\relax [ \left \{ 0 \right \},\left \{  \right \},\left \{  \right \} ] $ \\
\hline
$(2,1,1,0)$ & $\relax [ \left \{ 0 \right \},\left \{ 0 \right \},\left \{  \right \} ] $ \\
\hline
$(2,2,2,2)$ & $\relax [ \left \{  \sqrt{\frac{3}{20}}, - \sqrt{\frac{3}{20}} \right \},\left \{  \sqrt{\frac{3}{20}}, - \sqrt{\frac{3}{20}} \right \},\left \{  \sqrt{\frac{1}{5}}, - \sqrt{\frac{1}{5}} \right \} ] $ \\ \hline
$(4,1,0,0)$ & $\relax [ \left \{ 0 \right \},\left \{  \right \},\left \{  \right \} ] $ \\ \hline
$(4,1,1,0)$ & $\relax [ \left \{ 0 \right \},\left \{ 0 \right \},\left \{  \right \} ] $ \\ \hline
$(4,2,0,0)$ & $\relax [ \left \{ \frac{1}{2\sqrt{3}}, -\frac{1}{2\sqrt{3}} \right \},\left \{  \right \},\left \{  \right \} ] $ \\ \hline
$(4,2,0,1)$ & $\relax [ \left \{ \frac{1}{2}, -\frac{1}{2} \right \},\left \{  \right \},\left \{ 0 \right \} ] $ \\ \hline
$(4,2,1,0)$ & $\relax [ \left \{ \frac{1}{2\sqrt{3}}, -\frac{1}{2\sqrt{3}} \right \},\left \{ 0 \right \},\left \{  \right \} ] $ \\ \hline
$(4,2,2,0)$ & $\relax [ \left \{ \frac{1}{2\sqrt{3}}, -\frac{1}{2\sqrt{3}} \right \},\left \{ \frac{1}{2\sqrt{3}}, -\frac{1}{2\sqrt{3}} \right \},\left \{  \right \} ] $ \\ \hline
$(4,2,2,1)$ & $\relax [ \left \{ \frac{1}{2}, -\frac{1}{2} \right \},\left \{ \frac{1}{2}, -\frac{1}{2} \right \},\left \{ 0 \right \} ] $ \\ \hline
$(4,2,2,2)$ & \(
\begin{aligned}
 \relax [ &\left \{ -0.8096326613535235,0.8096326613535235 \right \},\\ &\left \{ -0.2109382698080428,0.2109382698080428 \right \}, \left \{ -\sqrt{\frac{1}{3}}, \sqrt{\frac{1}{3}} \right \}  ]
\end{aligned}
\) \\ \hline
$(4,3,1,2)$ & $\relax [ \left \{ \frac{1}{2}, -\frac{1}{2}, 0  \right \},\left \{ 0 \right \},\left \{ \sqrt{\frac{1}{6} },-\sqrt{\frac{1}{6} } \right \} ] $ \\ \hline
$(4,3,2,2)$ & \(
\begin{aligned}
 \relax [ &\left \{ -1.0078383823506314\ i,1.0078383823506314\ i,0 \right \},\\
&\left \{ -0.6773226457509125,0.6773226457509125 \right \},\\
&\left \{ -0.1687783139092837\ i,0.1687783139092837\ i \right \}  ]
\end{aligned}
\) \\ \hline
$(4,3,3,2)$ & \(
\begin{aligned}
 \relax [ &\left \{ -0.56102708691769753,0.56102708691769753,0 \right \},\\
&\left \{ -1.01831901409999213\ i,1.01831901409999213\ i,0 \right \},
\left \{ -\sqrt{\frac{1}{3}},\sqrt{\frac{1}{3}} \right \}  ]
\end{aligned}
\) \\ \hline
$(6,1,0,0)$ & $\relax [ \left \{ 0 \right \},\left \{  \right \},\left \{  \right \} ] $ \\ \hline
$(6,1,1,0)$ & $\relax [ \left \{ 0 \right \},\left \{ 0 \right \},\left \{  \right \} ] $ \\ \hline
$(6,1,2,0)$ & $\relax [ \left \{ 0 \right \},\left \{ \sqrt{\frac{5 \pm 2\sqrt{5}}{20}}, -\sqrt{\frac{5 \pm 2\sqrt{5}}{20}} \right \},\left \{  \right \} ] $ \\ \hline
$(6,1,3,0)$ & $\relax [ \left \{ 0 \right \},\left \{ \sqrt{\frac{2\sqrt{13} \pm 5 }{12}},-\sqrt{\frac{2\sqrt{13} \pm 5 }{12}} ,0\right \},\left \{  \right \} ] $ \\ \hline
$(6,1,3,2)$ & $\relax [ \left \{ 0 \right \},\left \{ \frac{1}{2\sqrt{3}},-\frac{1}{2\sqrt{3}}, 0\right \}, \left \{ \frac{1}{3},- \frac{1}{3}\right \} ] $ \\
            & $\relax [ \left \{ 0 \right \},\left \{ \frac{\sqrt{3}}{2},-\frac{\sqrt{3}}{2}, 0\right \}, \left \{ \frac{1}{\sqrt{3}},-\frac{1}{\sqrt{3}}\right \} ] $ \\ \hline
$(6,1,4,2)$ & \(
\begin{aligned}
 \relax [ &\left \{ 0 \right \}, \{ -0.6881909602355868,0.6881909602355868,\\
 &\ -0.1624598481164532,0.1624598481164532 \},\\
&\left \{ -0.7071067811865475,0.7071067811865475 \right \}  ]
\end{aligned}
\)
 \\ \hline
$(6,2,0,0)$ & $\relax [ \left \{  \sqrt{\frac{5 \pm 2\sqrt{5}}{20}},-\sqrt{\frac{5 \pm 2\sqrt{5}}{20}} \right \},\left \{  \right \},\left \{  \right \} ] $ \\ \hline
$(6,2,0,1)$ & $\relax [ \left \{ \frac{1}{2\sqrt{3} },-\frac{1}{2\sqrt{3} } \right \},\left \{  \right \},\left \{ 0 \right \} ] $ \\
            & $\relax [ \left \{ \frac{\sqrt{3}}{2 },-\frac{\sqrt{3}}{2 } \right \},\left \{  \right \},\left \{ 0 \right \} ] $ \\
\hline
$(6,2,2,0)$ & $\relax [ \left \{  \sqrt{\frac{5 \pm 2\sqrt{5}}{20}},-\sqrt{\frac{5 \pm 2\sqrt{5}}{20}} \right \},\left \{ \sqrt{\frac{5 \pm 2\sqrt{5}}{20}},-\sqrt{\frac{5 \pm 2\sqrt{5}}{20}} \right \},\left \{  \right \} ] $ \\
            & $\relax [ \left \{  \sqrt{\frac{5 \pm 2\sqrt{5}}{20}},-\sqrt{\frac{5 \pm 2\sqrt{5}}{20}} \right \},\left \{\sqrt{\frac{5 \mp 2\sqrt{5}}{20}},-\sqrt{\frac{5 \mp 2\sqrt{5}}{20}} \right \},\left \{  \right \} ] $ \\ \hline
$(6,2,2,1)$ & $\relax [ \left \{ \frac{1}{2\sqrt{3} },-\frac{1}{2\sqrt{3}}  \right \},\left \{ \frac{1}{2\sqrt{3} },-\frac{1}{2\sqrt{3}}  \right \},\left \{ 0 \right \} ] $\\
            & $\relax [ \left \{ \frac{1}{2\sqrt{3} },-\frac{1}{2\sqrt{3}}  \right \},\left \{ \frac{\sqrt{3} }{2},-\frac{\sqrt{3}}{2}  \right \},\left \{ 0 \right \} ] $ \\
            & $\relax [ \left \{ \frac{\sqrt{3} }{2},-\frac{\sqrt{3}}{2}  \right \},\left \{ \frac{1}{2\sqrt{3} },-\frac{1}{2\sqrt{3}}  \right \},\left \{ 0 \right \} ] $ \\
            & $\relax [ \left \{ \frac{\sqrt{3} }{2},-\frac{\sqrt{3}}{2}  \right \},\left \{  \frac{\sqrt{3} }{2},-\frac{\sqrt{3}}{2}  \right \},\left \{ 0 \right \} ] $ \\ \hline
$(6,2,2,2)$ & \(
\begin{aligned}
 \relax [ &\left \{ -0.1235329526036784,0.1235329526036784 \right \},\\
&\left \{ -0.1235329526036784,0.1235329526036784 \right \},\\
&\left \{ -0.3061580650652731,0.3061580650652731 \right \}  ]
\end{aligned}
\) \\ \hline
$(6,2,3,0)$ & $\relax [ \left \{ \sqrt{\frac{5 \pm 2\sqrt{5}}{20}},-\sqrt{\frac{5 \pm 2\sqrt{5}}{20}} \right \},\left \{ \sqrt{\frac{\pm 2 \sqrt{13}-5 }{12} },-\sqrt{\frac{\pm 2 \sqrt{13}-5 }{12} },0  \right \},\left \{  \right \} ] $ \\ \hline
$(6,3,0,0)$ & $\relax [ \left \{ \sqrt{\frac{\pm 2\sqrt{13}-5 }{12}},-\sqrt{\frac{\pm 2\sqrt{13}-5 }{12}},0  \right \},\left \{  \right \},\left \{  \right \} ] $ \\ \hline
$(6,3,2,2)$ & \(
\begin{aligned}
 \relax [ &\left \{ -1.0020126491889251\ i, 
 1.0020126491889251\ i, 0 \right \},\\
&\left \{ -1.1282504090231237, \
1.1282504090231237 \right \},\\
&\left \{ -0.6200482385775473, 0.6200482385775473 \right \}  ]
\end{aligned}
\) \\ \hline
$(6,3,3,0)$ & $\relax [ \left \{ \sqrt{\frac{\pm 2\sqrt{13}-5 }{12}},-\sqrt{\frac{\pm 2\sqrt{13}-5 }{12}},0  \right \},\left \{ \sqrt{\frac{\pm 2\sqrt{13}-5 }{12}},-\sqrt{\frac{\pm 2\sqrt{13}-5 }{12}},0 \right \},\left \{  \right \} ] $ \\ \hline
$(6,3,3,2)$ & \(
\begin{aligned}
 \relax [ &\left \{ -0.9990968994092514\ i, 
 0.9990968994092514\ i, 0 \right \},\\
&\left \{ -0.9990968994092514\ i, 
 0.9990968994092514\ i, 0 \right \},\\
&\left \{ -0.6692011970441518\ i, 
 0.6692011970441518\ i \right \}  ]
\end{aligned}
\) \\ \hline
$(6,4,0,2)$ & \(
\begin{aligned}
 \relax [ &\{ -0.1508691129768733,0.1508691129768733,\\
&\ -0.6008614236007921,0.6008614236007921 \}, \left \{  \right \},\\
&\left \{ -0.4659713220310156,0.4659713220310156 \right \}  ]
\end{aligned}
\) \\
            & \(
\begin{aligned}
 \relax [ &\{ -0.3804549074535447+0.5032996851609898\ i,\\
&\ \ 0.3804549074535447-0.5032996851609898\ i,\\
&\ -0.3804549074535447-0.5032996851609898\ i,\\
&\ \ 0.3804549074535447+0.5032996851609898\ i \}, \left \{  \right \},\\
&\left \{ -0.6195126630037510\ i,0.6195126630037510\ i \right \}  ]
\end{aligned}
\)
\\ \hline
$(6,4,2,2)$ & \(
\begin{aligned}
 \relax [ &\{ -1.0749698350912193\ i,1.0749698350912193\ i, \\
&\ -0.0764232867581285,0.0764232867581285 \},\\
&\left \{ -1.3183945773982646,1.3183945773982646 \right \},\\
&\left \{ -1.0785241468853727,1.0785241468853727 \right \}  ]
\end{aligned}
\) \\
            & \(
\begin{aligned}
 \relax [ &\{ -0.3961272181802985+0.5033454985998288\ i,\\
&\ \ 0.3961272181802985-0.5033454985998288\ i,\\
&\ -0.3961272181802985-0.5033454985998288\ i,\\
&\ \ 0.3961272181802985+0.5033454985998288\ i  \},\\
&\left \{ -0.9711343935705649,0.9711343935705649 \right \},\\
&\left \{ -0.5656662940384277\ i,0.5656662940384277\ i \right \}  ]
\end{aligned}
\) \\ 
            & \(
\begin{aligned}
 \relax [ &\{ -0.6981736132590754-0.4667135027424402\ i,\\
&\ \ 0.6981736132590754+0.4667135027424402\ i,\\
&\ \ 0.6981736132590754-0.4667135027424402\ i,\\
&\ -0.6981736132590754+0.4667135027424402\ i \},\\
&\left \{ -1.4637626767422020,1.4637626767422020 \right \},\\
&\left \{ -1.3351460192361050,1.3351460192361050 \right \}  ]
\end{aligned}
\) \\
\hline
$(8,2,0,0)$ & $\relax [ \left \{ 0.1141217371950750,-0.1141217371950750 \right \},\left \{  \right \},\left \{  \right \} ] $ \\ 
            & $\relax [ \left \{ 0.3987366944412020,-0.3987366944412020 \right \},\left \{  \right \},\left \{  \right \} ] $ \\ 
            & $\relax [ \left \{ 1.0382606982861683,-1.0382606982861683 \right \},\left \{  \right \},\left \{  \right \} ] $ \\ \hline
$(8,4,4,0)$ & \(
\begin{aligned}
 \relax [ &\{ 0.4632647275890309 + 0.5022938535699026\ i,\\
&\ -0.4632647275890309 - 0.5022938535699026\ i,\\
&\ \ 0.4632647275890309 - 0.5022938535699026\ i,\\
&\ -0.4632647275890309 + 0.5022938535699026\ i \},\\
&\{ 0.4632647275890309 + 0.5022938535699026\ i,\\
&\ -0.4632647275890309 - 0.5022938535699026\ i,\\
&\ \ 0.4632647275890309 - 0.5022938535699026\ i,\\
&\ -0.4632647275890309 + 0.5022938535699026\ i \},\\
&\left \{  \right \}  ]
\end{aligned}
\) \\ \hline
\end{longtable}

\end{document}